\documentclass[aps,prb,showpacs,twocolumn,floatfix,superscriptaddress]{revtex4-1}
\usepackage{graphicx,bbm,amssymb,amsfonts,amsmath,subfig,caption}  
\date{\today}

\begin{document}

\title{Length scales in the many-body localized phase and their spectral signatures}

\author{V.\ K.\ Varma}
\affiliation{Department of Physics and Astronomy, College of Staten Island, CUNY, Staten Island, NY 10314, USA}
\affiliation{Physics program and Initiative for the Theoretical Sciences, The Graduate Center, CUNY, New York, NY 10016, USA }
\affiliation{Department of Physics and Astronomy, University of Pittsburgh, Pittsburgh, PA 15260, USA}

\author{A.\ Raj}
\affiliation{Physics program and Initiative for the Theoretical Sciences, The Graduate Center, CUNY, New York, NY 10016, USA }

\author{S.\ Gopalakrishnan}
\affiliation{Department of Physics and Astronomy, College of Staten Island, CUNY, Staten Island, NY 10314, USA}
\affiliation{Physics program and Initiative for the Theoretical Sciences, The Graduate Center, CUNY, New York, NY 10016, USA }

\author{V.\ Oganesyan}
\affiliation{Department of Physics and Astronomy, College of Staten Island, CUNY, Staten Island, NY 10314, USA}
\affiliation{Physics program and Initiative for the Theoretical Sciences, The Graduate Center, CUNY, New York, NY 10016, USA }

\author{D.\ Pekker}
\affiliation{Department of Physics and Astronomy, University of Pittsburgh, Pittsburgh, PA 15260, USA}

\date{\today}

\vspace*{-1cm}

\begin{abstract}
We compute and compare the decay lengths of several correlation functions and effective coupling constants in the many-body localized (MBL) phase. 
To this end, we consider the distribution of the \emph{logarithms} of these couplings and correlators: in each case the log-coupling follows a normal distribution with mean and variance that grow linearly with separation. 
Thus, a localization length is asymptotically sharply defined for each of these quantities. These localization lengths differ numerically from one another, but all of them remain short up to the numerically observed MBL transition, 
indicating stability of the MBL phase against isolated ergodic inclusions.  
We also show how these broad distributions may be extracted using interferometric probes such as double electron-electron resonance (DEER) and the statistics of local spin precession frequencies.

\end{abstract}

\maketitle

\section{Introduction}
\label{sec:Intro}
Many-body localized (MBL) systems violate many of our expectations from equilibrium statistical mechanics: they do not thermalize under their own intrinsic dynamics 
\cite{BAA, Oganesyan:2007, ZnidaricPrelovsek, PalHuse}, have extensively many quasi-local conserved quantities \cite{lbitVadim, Abanin:2013,PekkerClark2017PRB},
 and retain the memory of their initial state at arbitrarily late times \cite{HuseReview}. 
These properties of MBL systems are stable to arbitrary (static) local perturbations; in this sense, MBL systems constitute a dynamical phase of matter, the properties of which have been extensively studied 
in the past decade \cite{ZnidaricPrelovsek, Bardarson, SerbynSlowGrowth, Schreiber2}. Most of these studies have considered one-dimensional systems, for which the MBL phase has been proved to exist under minimal assumptions \cite{Imbrie1, Imbrie2}.

The properties of the MBL phase are often characterized in terms of a localization length $\xi$. 
For instance, the growth of entanglement entropy starting from a product state follows\cite{ZnidaricPrelovsek, Bardarson, SerbynSlowGrowth} 
$S(t) \sim \xi \log t$; 
the high-temperature limit of the low-frequency a.c. conductivity is expected to behave as $\sigma(\omega) \sim \omega^{2 - \xi \log 2}$ for a spin-$1/2$ system \cite{SarangMarkus}; 
and an instability to isolated ergodic grains is believed\cite{deroeckHuveneers, Roeck2} to set in when $\xi = 1/\ln 2$. 
We have used the same symbol for these various quantities, but they are related to different correlation functions, and there is reason to doubt whether all correlations decay with the same $\xi$. 
Indeed, whether it makes sense to posit a well-defined $\xi$ in the MBL phase is unclear: based on numerical studies \cite{DavidVadim, Rademaker0, Rademaker, monthus2016flow, Pollet, PalSimon, ChandranKim}, 
analyses of rare-region effects \cite{KartiekReview}, and the structure of the locator perturbation theory \cite{Ros}, we expect all physical quantities in the MBL phase to exhibit strong fluctuations. 
\begin{figure}[h!]%
\includegraphics[width=1\columnwidth]{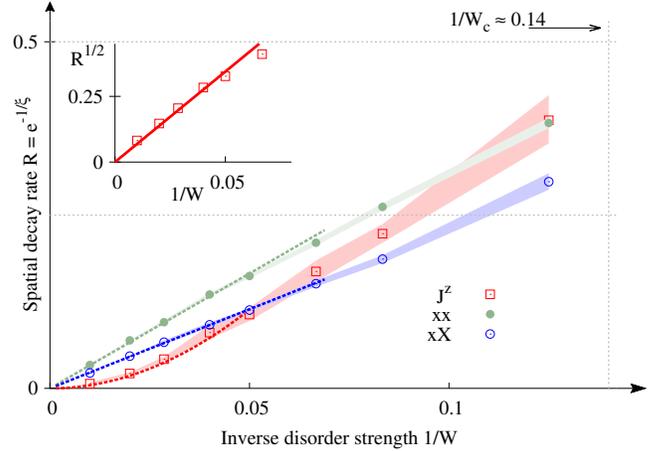}
\caption{ 
Spatial decay rates as a function of inverse disorder strength for effective coupling $J^z$ and correlators xx and xX in the MBL phase (see also Fig. \ref{fig:Jz-vs-xX} and Fig. \ref{fig:Jmed2}, MBL/ergodic transition $\frac{1}{W_c}\approx 0.14$). Dashed lines are fits to expected strong disorder behavior ($\frac{45}{W^{2}}, \frac{2.3}{W}, \frac{3.3}{W}$). Estimated uncertainty is indicated by shading. Gray dotted lines correspond to localization lengths $1/\log 2$ and $1/\log 4$, which are distinct estimates for the onset of the avalanche instability~(Sec.~3). 
Inset: Plot of $\sqrt{R}\sim 1/W$ for $J^z$ coupling.  %
    }%
    \label{fig:pxi}%
\end{figure}
The present work addresses and clarifies these issues, by analyzing how the probability distributions of correlation functions and effective coupling constants evolve with spatial separation in the MBL phase. 
These distributions are extracted numerically, using exact diagonalization and the Wegner-Wilson Flow (WWF) method. We find that in each case, the logarithm of the correlation function or coupling follows a 
normal distribution, with a mean and variance that grow linearly with spatial separation. 
These features are common to MBL and Anderson insulators; however, in the MBL case no simple relation seems to exist between the coefficients controlling the growth of the mean and variance. 
(By contrast, in Anderson insulators, for suitably defined quantities, the two coefficients are related by single-parameter scaling \cite{DMPK1, DMPK2, ATAF}.) %
The fact that variance only grows linearly in separation, $r$, guarantees that the inverse localization length for each coupling is sharply defined, with a distribution that narrows at large $r$. 
Thus one can ask how this quantity varies depending on the coupling. The corresponding spatial decay rate $R = \textrm{exp}(-1/\xi)$ exhibits simple dependencies on disorder strength anticipated by the structure of the perturbative "forward approximation"\cite{Ros}
(Fig. \ref{fig:pxi}). Importantly, all these lengths remain quite short inside the MBL phase previously identified in numerical studies. Our results suggest that this regime of disorder is also stable against rare 
disorder effects nucleating ergodic runaways \cite{deroeckHuveneers}.

Although the %
distribution for $1/\xi$ narrows for large separation, the couplings themselves become increasingly broadly distributed (log-normal), as noted above, with variance growing linearly in r. We develop protocols for extracting these broad distributions using an interferometric probe related to double-electron-electron resonance (DEER)~\cite{chuang_RMP, SarangPRL, kucsko_deer}. We show that a log-normal distribution of couplings implies that the disorder-averaged DEER response decays logarithmically in time, and also affects the statistics of local precession frequencies.

Sec. 2 sets up the notation, including model and methods, and also observables of interest. %
In Sec. 3 we present evidence for the ubiquity of broadening log-normal distributions that enable our definition of length scales shown in Fig. 1. %
Results on local spectra and spin echoes are presented in Secs. 4 and 5, respectively. We conclude with the discussion of the likely significance of our results and some future directions.
\section{definitions}
\label{sec:model}
\subsection{Model Hamiltonian: P-bits and L-bits}
We consider Heisenberg spin chains subject to random fields in the $z$ direction, described by the following Hamiltonian:
\begin{equation}
\label{eq:p-bit}
H=\sum_{i=1}^{L-1} \sigma_{i}^{x}\sigma_{i+1}^{x}+\sigma_{i}^{y}\sigma_{i+1}^{y}+\sigma_{i}^{z}\sigma_{i+1}^{z} + \sum_{i=1}^{L}h^{\phantom{z}}_i \sigma_i^z,
\end{equation}
where $\sigma_i^\alpha$ are Pauli matrices, $h_i$ is drawn from the distribution $h \in [-W, W]$, and the system size is $L$. 
We are interested mainly in the so-called Full MBL (FMBL) regime, where the entire spectrum of the system %
is MBL for $W\agt 10$ \cite{ZnidaricPrelovsek, PalHuse, Alet, Devakul} but will also include a value of $W=8$ in the transition regime.
Although the model conserves total $\sigma^z$ magnetisation and we may restrict ourselves to one with a fixed value of magnetisation, many of the correlators of interest mix magnetisation sectors; hence our results will be presented for the full model taking all sectors into account.

As described earlier, the FMBL phase possesses a complete set of ``local" L-bits\cite{lbitVadim} $[\tau^{z}_i,H]=0$ for $i \in [1,L]$ and corresponding Pauli raising and lowering operators $\tau^\pm_i$. 
The Hamiltonian can be re-written %
 as \cite{lbitVadim, Abanin:2013}
\begin{equation}
 \label{eq:lbit1}
\tilde{H} = E_0 + \sum_i B_i \tau_i^z + \sum_{i>j} J_{ij}\tau^z_i\tau^z_j + \sum_{i>j>k} J_{ijk}\tau^z_i\tau^z_j\tau^z_k + \ldots .
\end{equation}
The conserved charges $\tau^z$'s maybe thought of as obtained from $\sigma^z$ with appropriate dressing by (small) quantum fluctuations due to off-diagonal terms (in the z-basis)\cite{Ros}.
Multispin interactions $J_{ijk\ldots}$ induced by these virtual exchange processes should decay exponentially with the end-to-end distance among the spins (and therefore with the number of spins involved). In the regime where $W \gg 1$ we expect the dressing to be weak; therefore we expect the local fields $B_i$ to be close to the microscopic $h_i$, and the high-order terms in Eq.~\eqref{eq:lbit1} to fall off rapidly. 
\subsection{Observables}
\label{sec:observables}
\subsubsection{Couplings}
\label{sec:Jzs}
An alternative and useful representation of the L-bit Hamiltonian is obtained by focusing on subsystems of few spins treating the rest as a static environment, thereby trading infinitely many multispin interactions for \emph{distributions} of few spin terms. For example, if the subsystem is just a single L-bit at site $j$
\begin{equation}
H_{j|\text{env}}=B_{j|\text{env}}\tau^z_j,
\label{eq:H1}
\end{equation}
where the total number of $B_{j}$'s (and $H_{j}$'s) is $2^{L-1}$, with each rearrangement of environment's spins, $|\text{env}$, contributing to a spectral shift of an otherwise sharp local line. 
The statistics of these local fields is interesting and will be examined in Section \ref{sec:LSF}.  
If we instead look at two-spin subsystems with separation $r=j-k$ (henceforth we drop the subscript $|\text{env}$)
\begin{equation}
H_{j,k}=B_{j}\tau^z_j+B_{k}\tau^z_{k}+J^z_{j,k}\tau^z_{j}\tau^z_{k}
\label{eq:H2}
\end{equation} 
we can access the distribution of the two spin couplings, which is expected to exhibit a ``flow" previously described as an evolution to a broad $1/f$ law in the MBL phase\cite{DavidVadim} as $r\to \infty$. 
The flow reverses towards narrow (approximately Gaussian) distributions in the ergodic phase with the critical regime appearing as a family of non-flowing ``scale invariant" distributions\cite{DavidVadim}. 
Importantly, the bulk of previously computed results\cite{DavidVadim} did \emph{not} use the effective two spin distributions described here but rather a variant with only the fluctuations in the intervening region 
(i.e. only $r$ spins bookended by the two-spin subsystem) accounted for. While the prior choice was physically motivated, e.g. with distributions' cardinality growing $\sim 2^r$ as the more and more spins 
``mediate" the 2-spin coupling, we found empirically in this work that including \emph{all} spins (and thereby fixing the cardinality to $2^{L-2}$ for all r) significantly changes (reduces) finite size effects 
and vastly improves the overall quality of simple exponential fits, enabling for unambiguous extraction of length scales.
\subsubsection{Transverse correlators}
In addition to distributions of one- and two- L-bit terms in $H$
we will be interested in two transverse correlators, $\sigma^x_j\sigma^x_{k}$ and $\sigma^x_j\tau^x_{k}$, which we will refer to as xx and xX (note that XX is trivial by construction). Both of these correlators are interesting, albeit for different reasons: xx involves physical P-bits and can therefore be measured directly, 
while xX is important in studying effects of MBL subsystems weakly coupled to other degrees of freedom, 
e.g. the argument and the analysis of the instability due to isolated ergodic grains\cite{deroeckHuveneers} makes plausible assumptions about xX. 
\emph{Importantly}, in what follows we only consider the \emph{amplitude} of these correlators, i.e. averages of $|\langle n|\sigma^x_i \sigma^x_j|n\rangle|$. 
These are  \emph{not} instantaneous equilibrium observables, as they can only be extracted from the Edwards-Anderson type ``persistence in time" order parameter,  
$|\mathcal{O}|\equiv\sqrt{\lim_{T\to \infty} \int^T_0 dt \langle \mathcal{O}(t)\mathcal{O}(0)\rangle/T}$, e.g. with  $\mathcal{O}=\sigma^x_i \sigma^x_j$ -- 
this point has been extensively discussed and used in the context of eigenstate order in Hamiltonian problems 
\cite{PekkerPRX, SarangPRL, SarangFermi}. 
Finally, we note that all averaging in this paper is done by computing expectation values in eigenstate and then performing the equal weight (infinite temperature) Gibbs average and finally averaging over disorder realizations. 
\subsection{Wegner-Wilson flow}
A priori there is no unique way for constructing the dressed operators $\tau^z$; this is because the similarity transformation that diagonalizes $H$ needs to come with a single, consistent labeling scheme that assigns an L-bit label to each of the $2^L$ eigenstates of $H$ (out of the $2^L!$ possible labeling schemes).
However, a posteriori, physical constraints help sieve out a good method for the constructing this transformation: well-defined spatial locality of couplings $J$, and tightness of fields $B$ 
about the physical onsite potentials. 
By any definition of localization, these are reasonable requirements to be satisfied. The renormalization group-like technique of Wegner-Wilson flow (WWF) \cite{Wegner, Kehrein, DavidVadim} achieves these admirably.  In localized systems specifically, there are clear reasons for why WWF works well, as discussed in the literature -- this has to do with the order in which off-diagonal matrix elements are eliminated and the reversibility of the flow. 
\footnote{WWF can also be considered as an adiabatic flow between infinite and finite disorder where the invariant object is the entire set of bitstring labels of all many-body eigenstates -- these are conserved by definition under the flow and allow to interpret the results even outside the localized phase. Once defined, WWF continues to work across the phase diagram and allows for a seamless discussion of observables and their distributions -- that is very convenient, but may be confusing or ambiguous, e.g. if the L-bits are strongly smeared.}

Let the WWF parameter be labelled as $\kappa$; then the flow equation that transforms $H \rightarrow \tilde{H}$ is given by
\begin{equation}
 \label{eq:flow1}
 \frac{d H(\kappa)}{d \kappa} = \mathcal{G}(\kappa, H),
\end{equation}
where the function $\mathcal{G} := [H, \eta]$ is chosen so that (a) there is good stability in the flow as $\kappa : 0\rightarrow \infty$, and (b) the parts of the spectrum that we want to eliminate 
fall off quicker if they have larger energy separation.
A canonical choice of the generator $\eta$ \cite{Wegner, Kehrein}, in particular for diagonalising the entire system \cite{DavidVadim}, is given by
\begin{equation}
 \label{flow:Gen}
 \eta (\kappa) = [V(\kappa), H(\kappa)],
\end{equation}
where $V(\kappa)$ is the off-diagonal part of $H(\kappa)$, and $\eta(\kappa) = -\eta(\kappa)^{\dagger}$ is an antiunitary generator.

With these flow equations (and the understanding that $H(0) = H$ and $H(\infty) = \tilde{H}$) the transformation relating the two representations Eq. \eqref{eq:p-bit} and Eq. \eqref{eq:lbit1} is given by 
\begin{equation}
 \label{eq:Transform}
 \tilde{H} = U^{-1}HU,
\end{equation}
where columns of $U$ contain the L-bit representations of the eigenstates in the correct order as defined in the original \textit{P-bit} representation. 
Then the dressed spin operators are given by 
\begin{equation}
 \label{eq:STransform}
 \tau^{\alpha} = U \sigma^{\alpha} U^{-1},
\end{equation}
for $\alpha = x, y, z$
The transformation matrix $U$ that holds these L-bit representations is governed by a similar flow equation
\begin{equation}
 \label{eq:Uflow}
 \frac{d U(\kappa)}{d \kappa} = U \eta,
\end{equation}
with $U(0) = \mathbbm{1}$. Columns of $U$ along with the diagonal of $\tilde{H}$ completely characterises the spectrum of the problem.
\footnote{
A technical note is in order here: while taking the limit $\kappa \rightarrow \infty$ is not practically possible, we implemenent the flow using an adaptive 4,5 Dormand-Pince algorithm 
until the error estimate of $H(\kappa) \rightarrow H(\kappa + d\kappa)$ is below some threshold $-\log_{10} \epsilon = 3\sim 6$, concomitantly chopping away off-diagonal elements that are smaller than $n\epsilon$, 
where $n = \mathcal{O}(1) > 1$. The latter aspect is required in order to speed up the convergence to the effective Hamiltonian; 
$n<1$, on the other hand, will result in very minimal chopping away, and hence much slower convergence. As long as $n = \mathcal{O}(1)$, the chopping is not too aggressive.

Due to these approximations, the eigenvalues and eigenvectors are only approximate; they are matched to machine-precision eigenvalues and eigenvectors obtained from exact 
diagonalisation (ED) using the Hungarian matching algorithm \cite{Kuhn} i.e. bipartite graph matching for maximum sum of weights in the overlap of $U U_{ED}^{T}$. }
\section{Distributions and localisation lengths}
\label{sec:lengths}

\begin{figure*}[ttp!]%
    \centering
    \captionsetup{position=top}
    \raisebox{0.1cm}{\subfloat[W=25: decay]{{\includegraphics[height=4.5cm,width=6.cm]{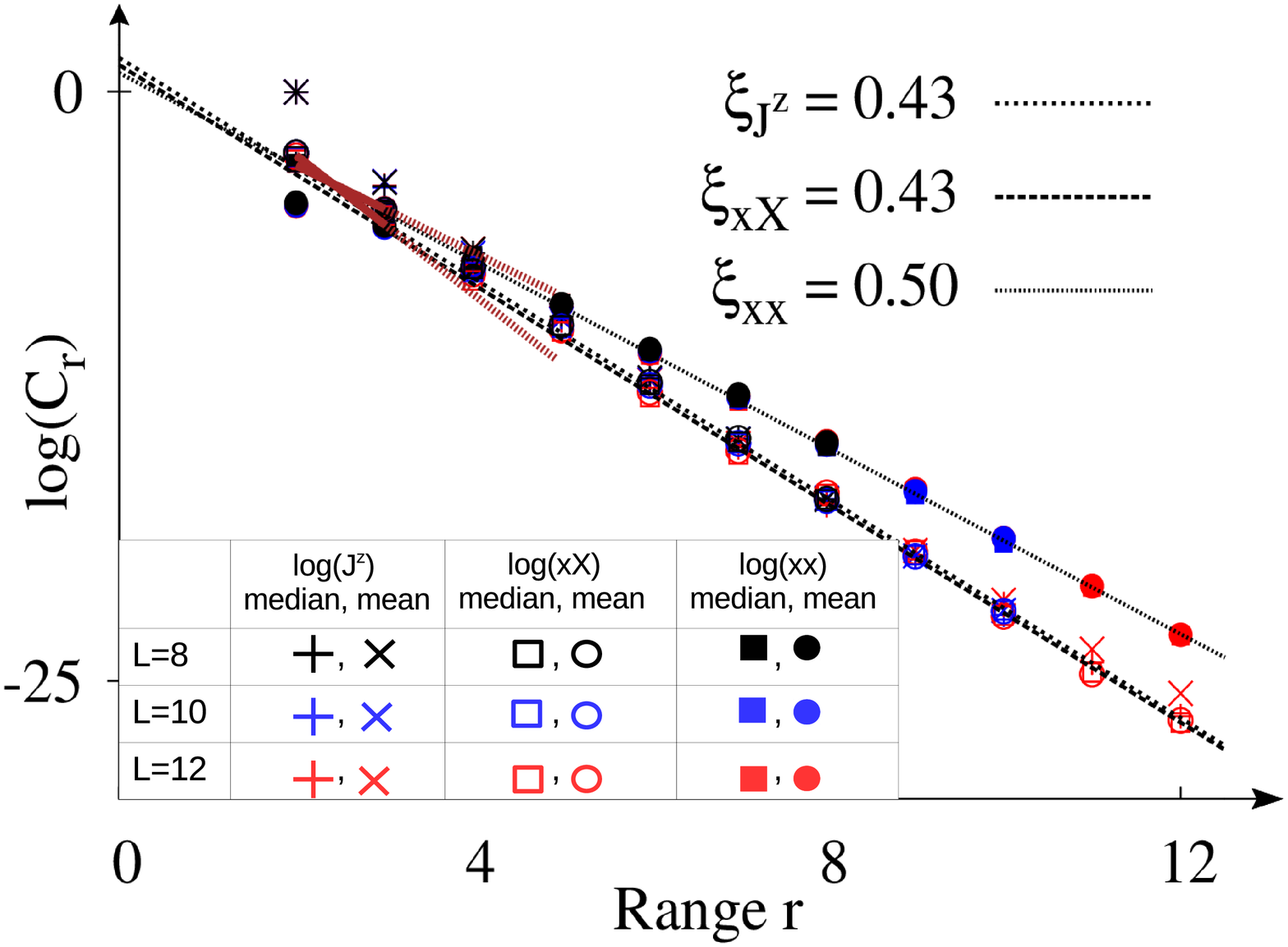} }}}%
    \subfloat[W=15: decay]{{\includegraphics[width=6.cm]{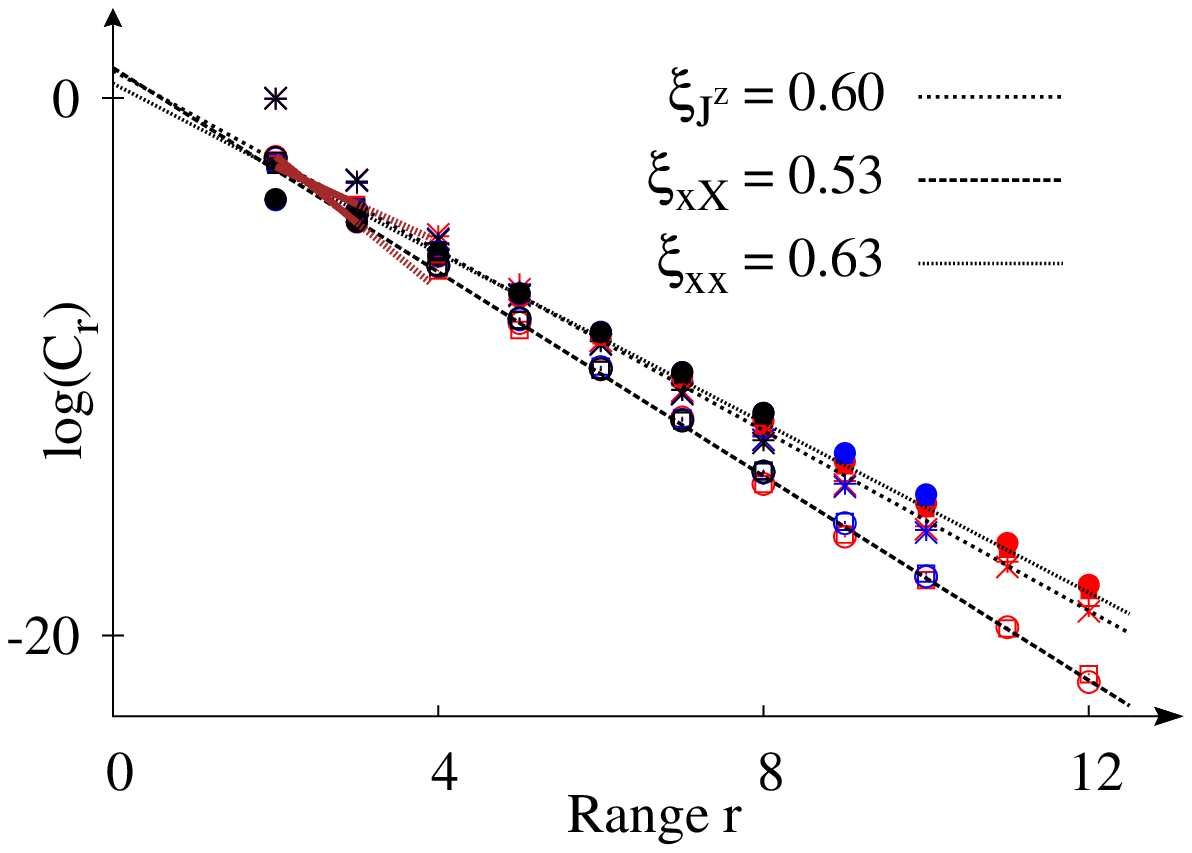} }}%
    \subfloat[W=8: decay]{{\includegraphics[width=6cm]{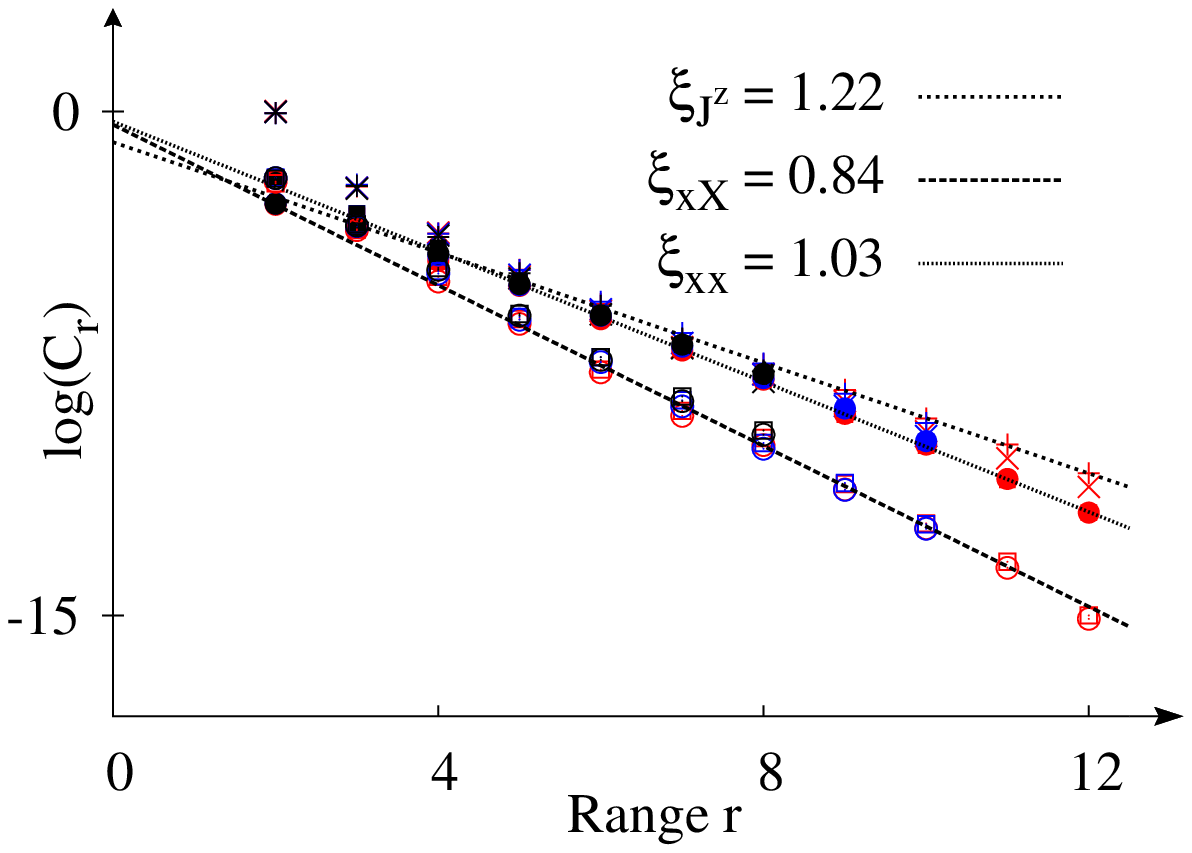} }}%
    \\
    \subfloat[$J^z$ collapse]{{\includegraphics[width=6cm]{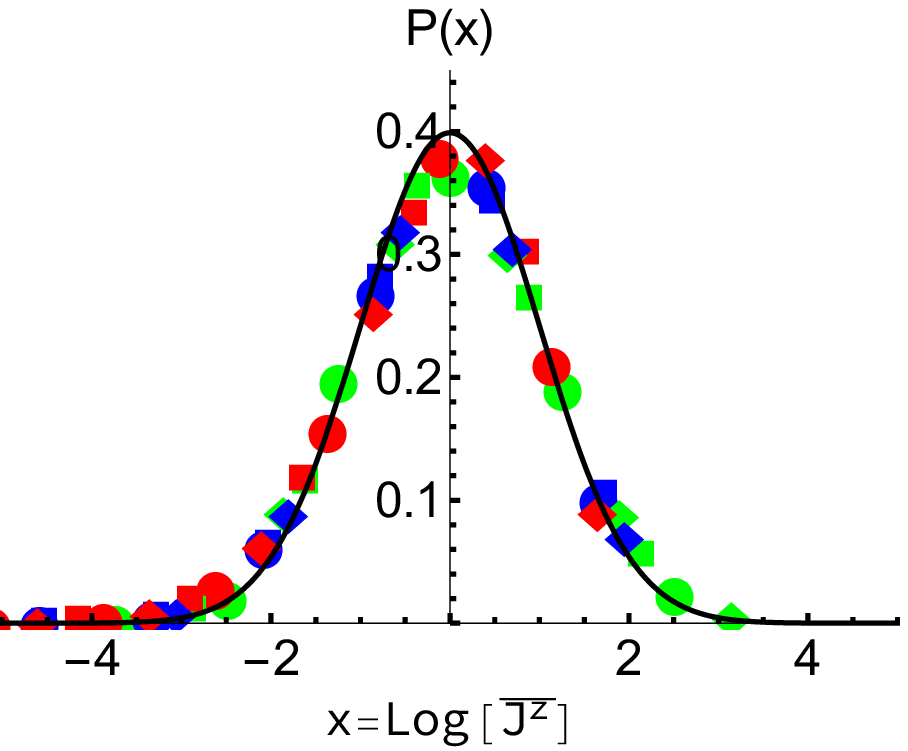} }}%
    \subfloat[xX collapse]{{\includegraphics[width=6cm]{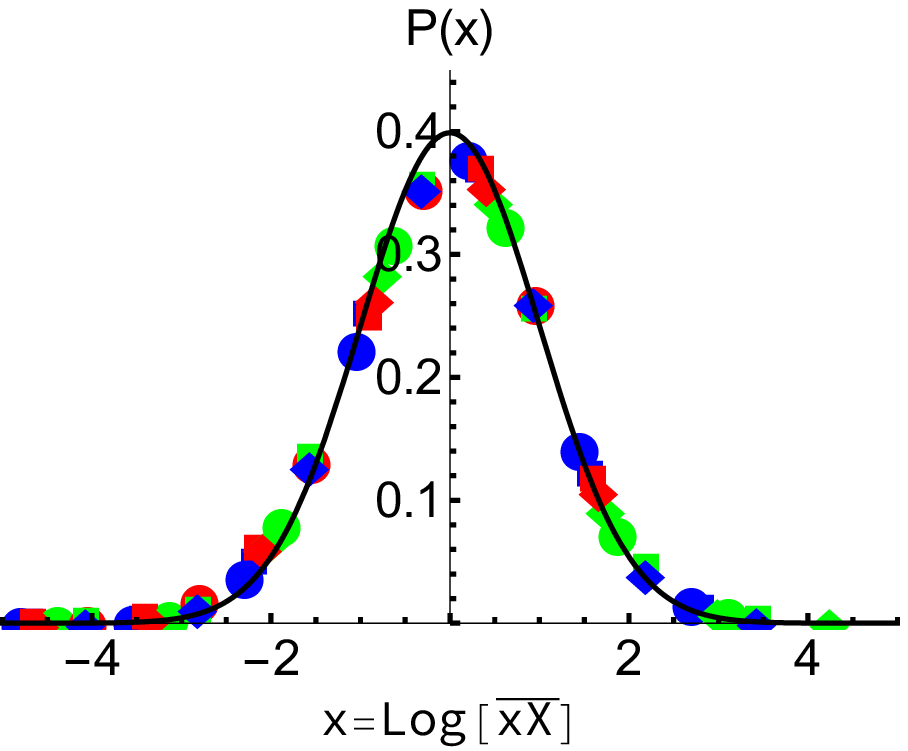} }}%
    \subfloat[xx collapse]{{\includegraphics[width=6cm]{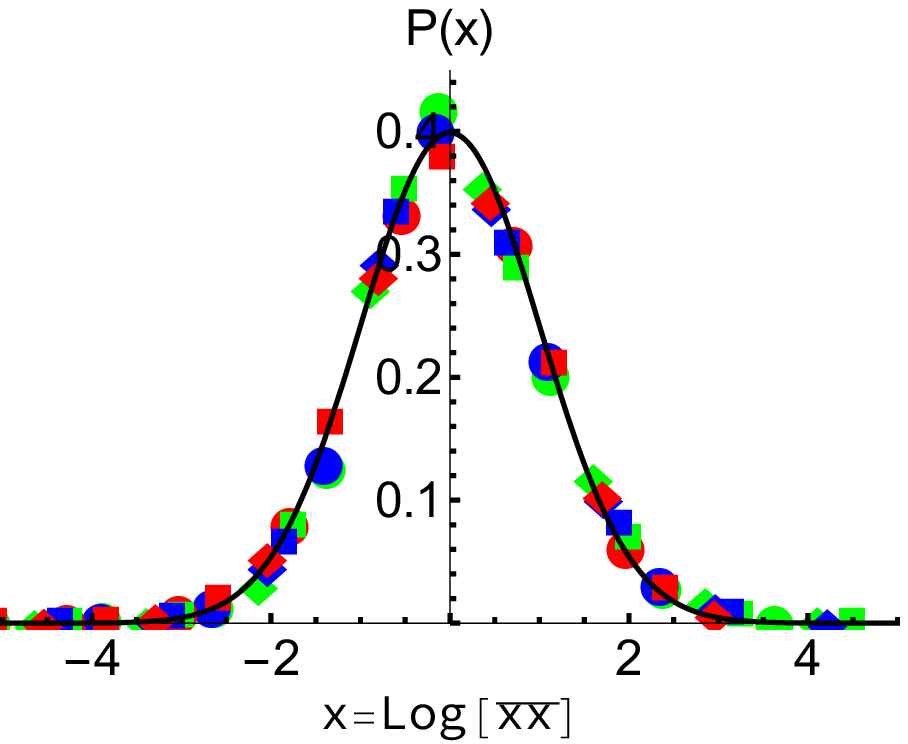} }}%
\caption{Log-normal statistics of $J^z, \langle\sigma^x\tau^x\rangle, \langle\sigma^x\sigma^x\rangle$ in $L=8,10,12$ (black, blue, red; color online) chains with $W=25$. 
(a)-(c) Decay with distance of the median of the logarithm of couplings at three disorder strengths, with a straight line on a semi-log scale indicating exponential dependence. Different system sizes overlap signaling insignificant finite size effects; the mean of logarithm (crosses, closed circles, open circles) are also shown which are largely indistinguishable 
from logarithm of median, except for $J^z$ where the crosses and pluses tend to deviate for largest $r$ (due to numerical underflow).
Perturbation theory, shown as thick brown lines at short distances, extrapolated beyond short distance with dashed brown lines, confirms the quantitative difference between xx and xX correlators.
The localization length fits (thin lines) from fitting the last 60\% of the decays of the three couplings in each panel are indicated in the legends.
(d)-(f) Distribution of the three couplings; overbar symbols refer to the data being normalized by its standard deviation and median, so as to achieve data collapse across various ranges 
$r = 6 (\textrm{circle}), 8 (\textrm{square}), 10 (\textrm{diamond})$ and 
disorder strengths $W = 25 (\textrm{green}), 15 (\textrm{blue}), 8 (\textrm{red})$; system size is fixed at $L=10$ here. The black line is a normal distribution with zero mean and unit variance.}
    \label{fig:Jz-vs-xX}%
\end{figure*}
This section makes two essential points. First, log-normal distributions are natural and pervasive in localized problems. Second, unlike in Anderson localized (single particle) case, multiple length scales are required to characterize distributions in many-body problems.
\subsection{Warmup -- single particle and non-interacting fermions}
It is well known that the wavefunction intensity $G=|\psi|^2$ (or measurable conductance) of Anderson localized single particle states in 1D is log-normally distributed \cite{DMPK1, DMPK2, ATAF}
\begin{equation}
 \label{eq:DMPK}
 P(\log G) = \sqrt{\frac{\xi_2}{8\pi r}}\textrm{exp}\left[ -\frac{(\log G + 2r/\xi_1)^2}{8 r /\xi_2}\right],
\end{equation}
with $r$ denoting spatial separation. Remarkably, both the mean and variance of $G$ are controlled by the same $\xi=\xi_1=\xi_2$ -- the localization length -- 
that fully characterises the localised phase\cite{ATAF,DMPK1,DMPK2}. 
That this length (or rather its inverse) is sharply defined is made clear by considering a distribution for $1/\xi\equiv-\log G/(2 r)$ which narrows to a $\delta$ function in the limit $r\to\infty$. 
The fluctuations in $1/\xi$ remain important for some observable quantitites, e.g. they are known to alter the prefactor to the Mott a.c. conductivity of disordered systems \cite{MottHybrid}. %
Importantly, $\xi$ in Eq. \ref{eq:DMPK} varies somewhat with energy even at strong disorder and appreciably at weak disorder. This, combined with specifics of averaging, may result in significantly different statistics in non-interacting many-body problems, e.g. at low but finite temperature in a weakly interacting regime.

With an eye to interacting spin chains, we consider infinite-temperature correlators in the noninteracting many-fermion problem in order to understand the distinction between xx and xX correlators.
We define locator fermion annihilation operator on site $j$, $c_j$, and also that of the unique eigenstate ``attached" to site $j$ in perturbation theory $\gamma_j$. The point we wish to make is that xx and xX spin correlators defined above are analogous (sans Jordan-Wigner phase) to 
\begin{align}
\langle n| c_i^+ c_{j} |n \rangle &= \sum_k f(k) \psi^*_k(i) \psi_k(j)
\label{eq:fermxx}
\\
\langle n|  c_i^+ \gamma_{j}|n \rangle&=f(j)\psi^*_j(i),
\label{eq:fermxX}
\end{align}
respectively. (Note: $f(\alpha)=0,1$ is the occupation of each single-particle state $\alpha$ in $|n\rangle$) Comparing typical decay rates of the two correlators we expect the first one to decay slower since it involves a sum over many orbitals and is dominated by the slowest-decaying orbitals that contribute.
Fluctuations in the single-particle localization length, and particularly its energy-dependence, imply that the length-scale in Eqs. \eqref{eq:fermxx} should systematically exceed that in Eq. \eqref{eq:fermxX}.%

\subsection{Many-body case}    
We now present one of the central results: in Fig. \ref{fig:Jz-vs-xX} -- the distributions of xx and xX correlators and of $J^z_{i,j}$ effective couplings sampled over many-body eigenstates and disorder realizations. 
Their apparently log-normal shape and linear growth (with separation) of both mean-log and var-log naturally leads to a spectrum of length scale parameters, two per observable.
Following the non-interacting example we expect xx to decay slower than xX and also display weaker fluctuations, as it ``pre-averages" over single particle fluctuations.  
These expectations are clearly borne out (see Figs. \ref{fig:pxi}, \ref{fig:Jz-vs-xX} for averaged quantities, and Fig. \ref{fig:SingleSample} for a single sample.). 
Finally, we expect and observe that the spatial decay rate of the typical $J^z$ is $e^{-1/\xi_1}\sim 1/W^2$. At strong disorder, the leading contribution to the typical $J^z$ comes from Hartree shifts of the nearly site-localized orbitals. These Hartree shifts induced by site $i$ at site $j$ scale as the \emph{square} of the amplitude of orbital $i$ at site $j$; since the amplitudes decay as $(1/W)^{|j - i|}$, the effective interaction $J^z$ decays as $(1/W^2)^{|i - j|}$. 
This difference in dependence on $1/W$ is borne out as seen in Fig. \ref{fig:pxi}; the error bars are determined by the uncertainties of fitting various segments and various system sizes. 

Two comments are in order. First, all length scales are quite short, suggesting that most of the numerically observed localized phase (for $W \agt 10$, corresponding to $W \agt 5$ in the notation of Ref.~\cite{PalHuse}) is stable against the inclusion of ergodic grains.
The exact threshold for the critical localization length in the generic interacting case is not fully settled. In the case of noninteracting l-bits~\cite{Roeck2}, the instability happens when $\xi = 1/\log 2$. When the l-bits are interacting, a given l-bit at a distance $r$ from the ergodic inclusion can couple to the inclusion via many distinct processes~\footnote{D. A. Huse, private communication} (depending on whether the $r - 1$ intermediate l-bits get flipped or not) and a conservative estimate of when the instability sets in is $\xi = 1/\log 4$. Both criteria are marked in Fig.~\ref{fig:pxi}. Second, much of the ensemble-averaged physics is present in single samples as well; see Fig. \ref{fig:SingleSample} where we present the decays of the three observables for two different samples with different 
disparate $\xi$ values, one small and one big. We find that the xx and xX decays still follow the same trend with respect to each other, in particular Eq. \eqref{eq:xxVsxX}, as we see from the middle column of plots. 
Moreover the right column of Fig. \ref{fig:SingleSample} shows, upon comparing with its middle column, that quicker decays of couplings (stronger localization) results in 
a broader spread of local field splittings (see next section for its definitions) and hence broader spread of $J^z$ couplings.

To summarize, each two-point object of interest may be efficiently ``labeled" using two length scales, $\xi_1$ and $\xi_2$, encoding typical decay and growth of fluctuations. The first of these connects to the established body of work where properties of the MBL phase were understood using L-bit phenomenology, albeit with a potentially important caveat that different observables are controlled by numerically different decay lengths. It is an interesting exercise to re-examine and correct these prior results to account for fluctuations. In some cases we expect the corrected answer to be qualitatively similar to the mean-field one. 
For example, the logarithmic growth of entanglement will now come with strong but subdominant fluctuations -- the so called ``logarithmic lightcone" will be smoothed out on scales $\sim \sqrt{\log t}$, i.e. there is a broad but still discernible boundary demarcating entangled spins in real space. Unless the two lengths turn out vastly different (they are not in our case but may be in other models) the mean-field theory is still a reasonable if incomplete description in such cases.
The more interesting possibility, perhaps (of course!), is to find examples of phenomena where these fluctuations invalidate mean-field expectations entirely -- we focus on two such cases next.

\section{Local spectral functions}
\label{sec:LSF}
\begin{figure*}%
    \centering
    \captionsetup{position=top}
    \subfloat[Difference of local fields: WWF]{{\includegraphics[width=8.5cm]{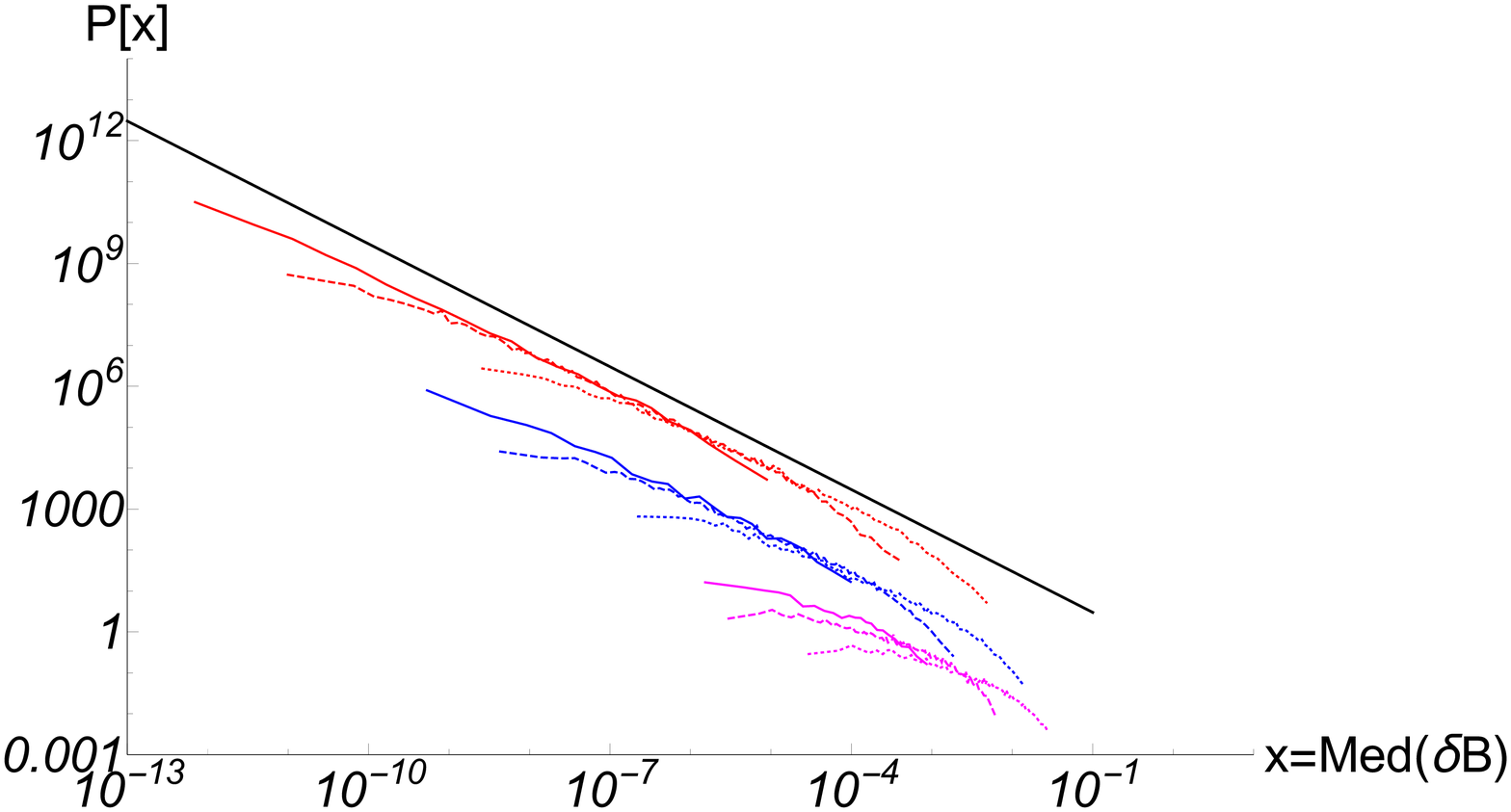} }}%
    \raisebox{0.05cm}{\subfloat[Spectral tree]{{\includegraphics[height = 4.8cm, width=8.5cm]{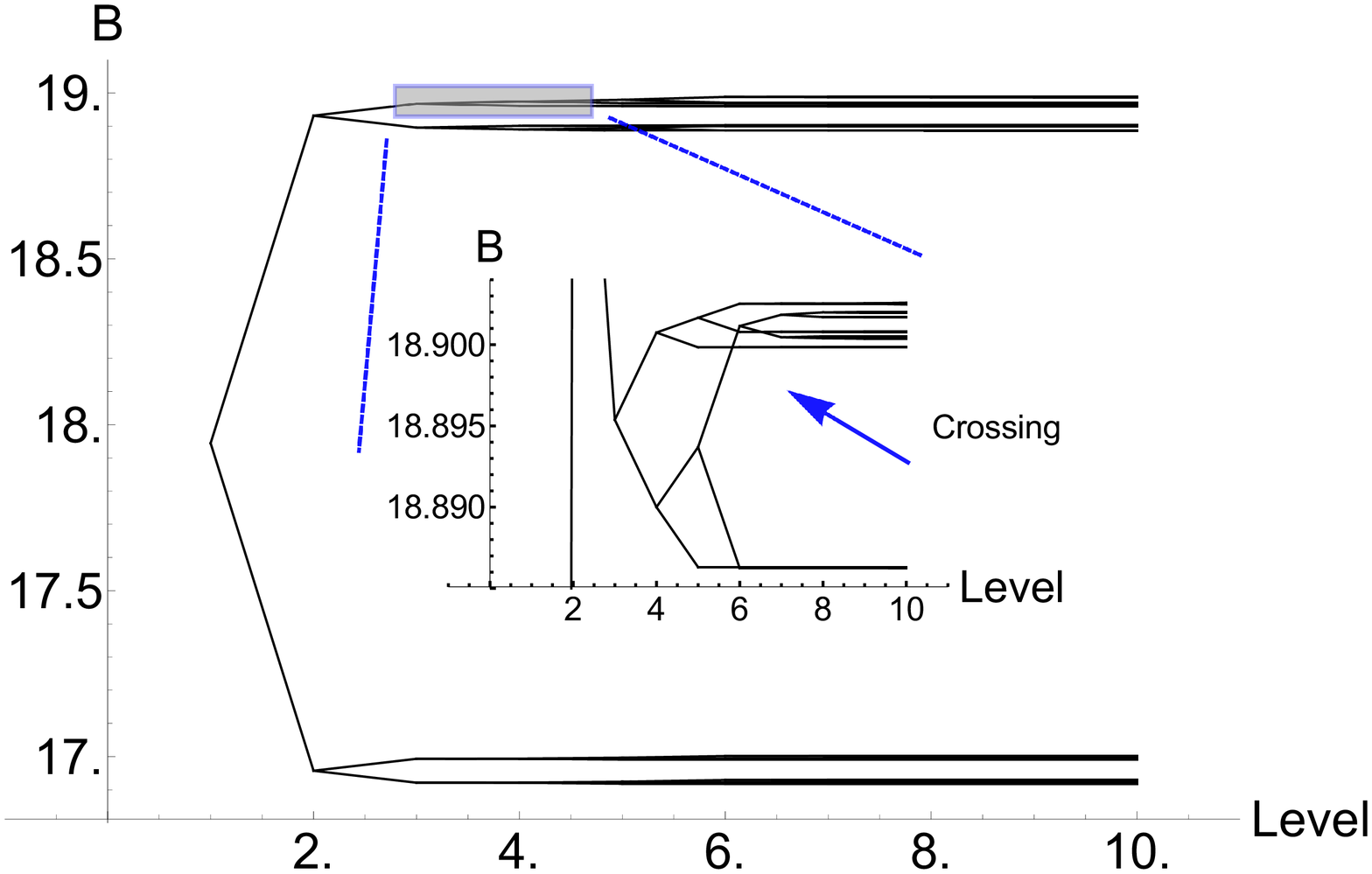} }}}%
    \\
     \raisebox{0.0cm}{\subfloat[Hamming distance]{{\includegraphics[height = 4.8cm, width=8.5cm]{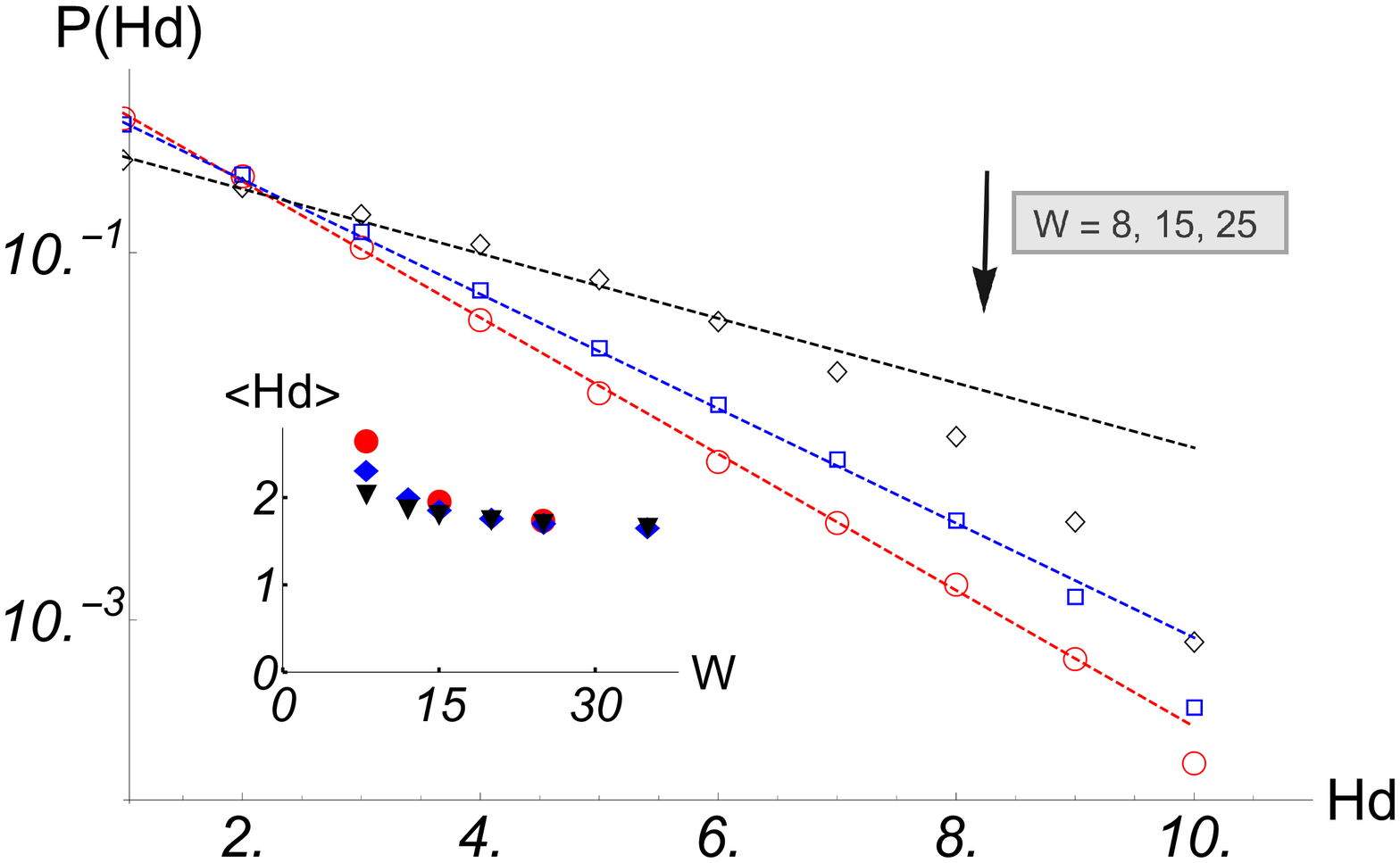} }}}%
    \subfloat[Difference of local fields: Mean field results $\xi = 1.0$]{{\includegraphics[height = 4.8cm, width=8.5cm]{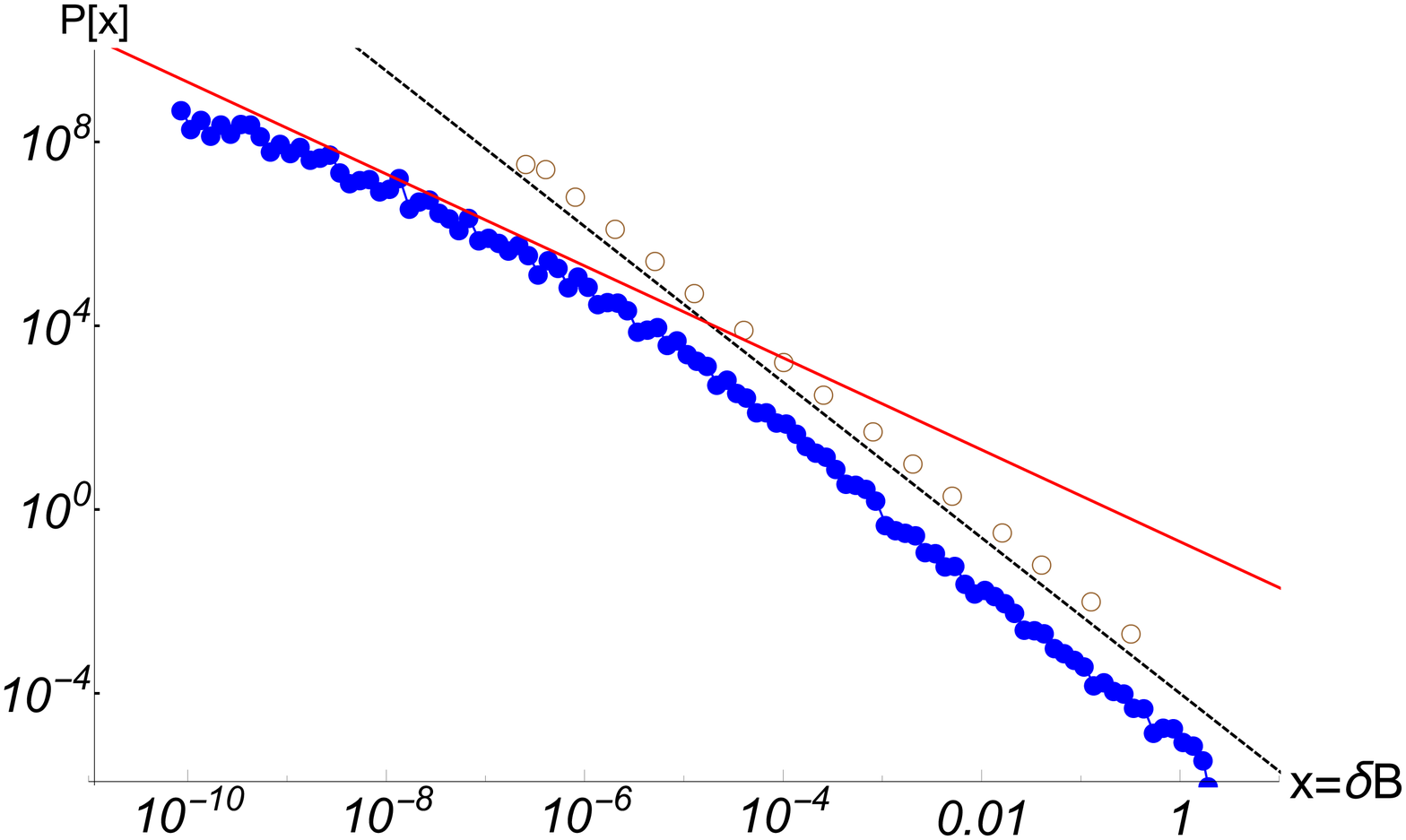} }}%
    \caption{
    (a) Spectral shift statistics for $W = 25, 15, 8$ (red, blue, magenta; vertically shifted for clarity) in the localized phase; for a fixed disorder strength, the three curves (dotted, dashed, full) correspond to increasing range ($L=8,10,12$).
     (b) Spectral tree showing absolute values of the fields and the splittings for a single sample with small $\xi$ . 
    A larger $\xi$ of the exponential decay of couplings will result in more crossings in the sample's tree.
     (c) Distribution of Hamming distances for $L=12$ chain for various disorder strengths, $W=25,15,8$ (red/circle, blue/square, black/diamond, respectively).
     The dashed lines of the same color show the fits of the respective symbols to an exponential distribution, which become progressively better as $W$ increases.
     Inset: mean Hamming distances as a function of $W$ for various $L = 8,10,12$ (red/circle, diamond/blue, triangle/black), which unambiguously demonstrate its decrease as $W$ increases.
    (d) Mean field tree distributions of spectral shifts %
		$P(\delta B)$ from $L=18$ chain of mean-field couplings for localisation length $\xi=1$, with open-brown being purely mean-field and closed-blue including 
		fluctuations in couplings $\textrm{exp}(-r/\xi + \sqrt{r}\delta)$, with $\delta = [-2,2]$ drawn randomly. 
		The expected %
		$P(\delta B) \sim (\delta B)^{-1 - \xi \log{2}}$ is shown by the dashed-black line, and $1/x$ power-law is shown by red-full line.  }
    \label{fig:SpectralProps}%
\end{figure*}
Local spectra can be defined and measured experimentally via autocorrelations of single spin operators, e.g. Gibbs-averaged 
\begin{equation}
A_{\textrm{xx}}(\omega)=\int dt e^{i \omega t} \langle \sigma^x_j (t)\sigma^x_j (0)\rangle.
\label{eq:Axx}
\end{equation}
These are generally complicated convolutions of spectra and matrix elements. For a chain of $L$ spins we expect $\sim 2^{2L}$ broadly distributed contributions to $A_{\textrm{xx}}(\omega)$.  In MBL systems, with well- (or at least usefully) defined L-bits, $\sim2^L$ of these values are parametrically larger than the rest. We can vastly simplify (``clean up") the situation by considering spectral functions of L-bits 
(its relationship to $A_{\textrm{xx}}(\omega)$ will be explained at the end of this Section)
\begin{equation}
A_{\textrm{XX}}(\omega)=\int dt e^{i \omega t} \langle \tau^x_j (t)\tau^x_j (0)\rangle,
\label{eq:AXX}
\end{equation}
thereby removing the fluctuations of the matrix elements and vastly reducing the number of terms, down to $2^L$, with $A_{\textrm{XX}}$ equal to the probability distribution of local fields in Eq. \ref{eq:H1}, 
$P(B_{j})$. The spectral line starts infinitely sharp in non-interacting systems and becomes splintered by L-bit interactions. %
 We will focus on the distribution of \emph{spectral shifts} (i.e., spacings between adjacent frequencies), $P(\delta B_{j}=B_{j+1}-B_{j})$, to elucidate local spectral correlations. 

We find that $P(\delta B)$ appears to follow log-normal statistics with a dramatic overall increase in the dynamical range upon entering the MBL phase (see Fig. \ref{fig:SpectralProps}(a)) implying a simple $1/f$ powerlaw as before, albeit with system size $L$ providing regularization instead of separation $r$ in two point observables. We argue below that to reproduce this power law one must include fluctuations of the $J^z$ couplings.

To start, it is helpful to visualize the local spectrum as a splintering process using a spectral tree\cite{NGH}, see Fig. \ref{fig:SpectralProps}(b). 
Here the root of the tree is the frequency of the isolated site $j$ (average of $B_{j}$ over all the environment spins, corresponding to the local onsite potential), 
and each generation corresponds to incrementally turning on (exponentially decaying) interactions to further neighbors (or equivalently, only averaging interactions with progressively distant spins). 
Since each site has a spectral tree associated with it, we have chosen the rightmost boundary site from each sample (so that the starting value in each tree is approximately the onsite potential at $x=10$); 
note that the boundary sites show the strongest splitting because of the largest available distances.
Deep inside the MBL phase we expect a very rapid decay of the coupling and the tree not to cross itself, with half of $\delta B$'s \emph{equal} to (twice) some interaction term with the most distant spin, 
a quarter of $\delta B$'s corresponding to sums and differences of two interaction terms etc.
Put differently, there is a considerable ``which path" or ``branching" memory which should manifest in a fractal-like structure of local MBL spectra and a non-trivial non-universal powerlaw distribution 
$P(\delta B)\sim |\delta B|^{-1-\xi_1 \log 2}$ (see appendix). 
By contrast, when interactions are strong, we do not expect any branching memory, with each $\delta B$ obtained from a random combination of many $J$'s -- adjacent frequencies correspond to configurations 
that differ by several spin-flips, hence $P(\delta B)$ should obey Poisson statistics in the ergodic phase. 
This transition takes place, within the mean-field picture, at $e^{-1/\xi_1}=1/2$ (note that here we are considering the special case of an edge spin~\cite{NGH}). %
The degree of self-crossing in the spectral tree can be quantified, in fact, if we examine the distribution of Hamming distances corresponding to each $\delta B$ 
(see Appendix for details on its computation). 
These appear quite short in the MBL phase (see Fig. \ref{fig:SpectralProps}(c)), and is smaller than the fully ordered mean-field prediction of $\langle \textrm{Hd} \rangle = 2$, which can occur due to 
rearrangements of L-bits.
The fully-ordered mean-field case is easily seen to have a distribution of $P_{\textrm{MF}}(\text{Hd}) = 2^{-\text{Hd}}$. Away from this fully-ordered mean-field limit, we observe (Fig. \ref{fig:SpectralProps}(c)) that the actual data exhibits 
a generalized exponential behavior
\begin{equation}
 \label{eq:PHd}
 P(\text{Hd}) = (\kappa-1) \kappa^{-\text{Hd}},
\end{equation}
with $\kappa = \kappa(W)$, a nondecreasing function of disorder strength. Note that Eq. \eqref{eq:PHd} is a valid probability distribution defined at the positive integers, 
with $\langle \text{Hd} \rangle = \frac{\kappa}{\kappa-1}$. 
We therefore observe that while the tree remains essentially non-crossing as anticipated by the mean-field model, the spectral statistics are much more universal, with a simple powerlaw of $-1$. 

To understand this result we now allow multiplicative disorder in the coupling strength, to mimique log-normal distributions with growing variance (as detailed in the previous section). 
Thusly modified L-bit description may be simulated straightforwardly numerically, see Fig. \ref{fig:SpectralProps}(d). We clearly observe the existence of a simple $1/f$ powerlaw. 
Indeed, in principle, there should also be a crossover from a non-universal to the asymptotic $1/f$ powerlaw (at smallest $\delta B$'s) -- this requires $\xi_2/\xi_1 \ll 1$ 
which is not the case for spin chains studied here.

While the local L-bit spectral function is not directly measurable, when $\xi_1 < 1$ one can approximate it well by taking a physical spectral function, measured by standard spectroscopic means, 
and dropping all spectral lines below a certain threshold intensity when computing gaps. 
In practice, extremely small splittings will not be resolvable, so one can only measure the ``tree'' out to a depth set by experimental resolution. 
Assuming that the experimental resolution is $500$ times the microscopic energy scales (as is reasonable for present-day experiments with ultracold atoms and superconducting qubits \cite{Google1, Google2}), 
and that $1/\xi_1 = 2/3$, one can resolve up to four generations of the tree, which should be adequate to test the predicted hierarchy of gaps. 
The protocol to measure the energy spectrum of a set of spins or qubits consists of creating local excitations (through, say, a magnetic $\pi/2$ pulse) and measuring the time-dependent vibrational response: 
a simple Fourier transform will then reveal the characteristic modes (eigenenergies) of the system \cite{Google1}. 
In order to construct the tree, and the corresponding splittings, up to some desired level, the strongest $2^L$ spectral lines may be retained.

\section{Echoes and L-bit dynamics}
\label{sec:echoes}
\begin{figure}[tp]%
\includegraphics[width=0.99\columnwidth]{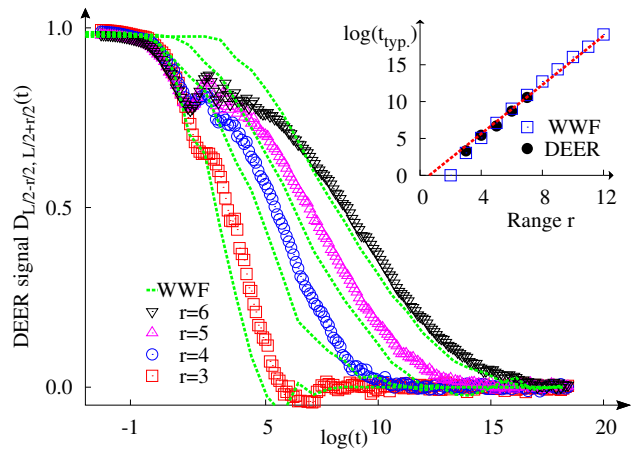}
\caption{ 
P-bit DEER signal for $W=15$ and $L=8$ compared against Fourier transform of the distribution of L-bit $J^z$ (Sec. \ref{sec:Jzs}). We rescaled the former signal by a small factor to line-up the traces at short-test times.
    }%
    \label{fig:pxiDEER}%
\end{figure}

The existence of L-bits may be demonstrated using several related but inequivalent dynamical protocols \cite{SarangPRL, Rademaker, VasseurMoore, ChandranKim}. 
The one of interest here is a \emph{local} Hahn echo which consists of a familiar (e.g. from NMR) sequence of pulses, however with all three pulses applied on the same site $j$ of the chain
\begin{equation}
|t\rangle=R_j^{\pi/2} e^{-iHt/2} R_j^{\pi} e^{-iHt/2} R_j^{\pi/2} %
|\uparrow\rangle,
\label{eq:Hahnt}
\end{equation}
where $R_j^{\phi}=\exp(i \phi \sigma_y/2)$
and $|\uparrow\rangle$ is a state with site $j$ fully polarized (either via quench or by pre-measurement.) 
With these manipulations the persistent echo is obtained in the MBL phase
\begin{equation}
D_j(t\to \infty)\equiv\langle t|  \sigma_j^z | t \rangle>0
\label{eq:Hahn}
\end{equation} 
The initial state may be a unique state, e.g. a particular product state that is easy to prepare or an eigenstate which might be subsequently averaged over to simulate a thermal distribution (which may be imposed by coupling weakly to the environment).
The only difference with textbook NMR discussion of Hahn echo is that there the pulses apply globally and can only rephase the decoherence from chemical shifts. Applying pulses locally in MBL systems effectively performs Hahn rephasing in the total static field comprised of local single body and interaction components. 
As previously discussed\cite{SarangPRL} the existence of the echo may be interpreted similarly to a finite quasiparticle residue in Fermi liquids which guarantees that calculations done in the renormalized 
model are directly measureable by coupling to actual degrees of freedom. 
MBL states in finite chains even at moderate disorder tend to support a reasonably visible echoe amplitude, e.g. at $W=8$ typical $D_j$ is about 0.5 (it is 1 for the perfect echo performed with exact L-bit rotations).

Building on this we design a two-spin echo [similar to the double electron-electron resonance (DEER)] to extract the information about the distribution of couplings $J_{jk}$ between two L-bits at a specified distance $r=|j-k|$. To start we ignore the difference between L-bits and P-bits, i.e. assume $D_j=1$. In this protocol, one performs a local Hahn spin echo on spin $j$; however, simultaneous with the $\pi$-pulse applied to spin $j$ in this protocol, one also applies a $\pi$-pulse to spin $k$
\begin{equation}
|t\rangle=R_j^{\pi/2} e^{-iHt/2} R_j^{\pi} R_k^{\pi} e^{-iHt/2} R_j^{\pi/2} |\uparrow\rangle.
\label{eq:Deer}
\end{equation}
All couplings acting on $j$ \emph{except} that due to $k$ remain echoed out as above, allowing one to measure $J_{jk}$ without having it masked by the stronger couplings due to spins closer to $j$ which allows for the signal to decohere in time following 
\begin{equation}
D_{jk}(t)=\langle e^{i J_{jk} t}\rangle=\int dJ P_{jk}(J) e^{i J t}.
\label{eq:ftdeer}
\end{equation}
The time-dependence of the \emph{averaged} DEER response is precisely the Fourier transform (i.e., characteristic function) of the probability distribution of $J_{jk}$ so the DEER protocol allows for a concrete test of our predictions concerning the distribution functions.
What should we expect for $[\mathcal{D}_{jk}(t)]$ assuming a log-normal distribution of $J$? 
We expect that $[\mathcal{D}_{jk}(t)]$ decays from nearly 1 to zero albeit logarithmically slowly, as implied by the $1/J$ prefactor in the log-normal distribution. 
It is especially illuminating to consider the dependence of the decay profile on the separation between the two spins in the echo $r=|j-k|$ -- signal's half-life, $t_{1/2}$, 
is directly determined by the typical (log-mean) coupling at that separation ($\log t_{1/2}\approx \log 1/J_{typ}$), while the log-slope reflects the fluctuations - it decreases as inverse root variance 
\begin{equation}
D_{jk}(t)\propto -\sqrt\frac{\xi_2}{r}\log t
\end{equation}
Our results at moderate disorder $W=15$ (Fig. \ref{fig:pxiDEER}) are clearly consistent with these expectations.
\section{Discussion} 
In this work we explored multiple ways of characterizing localization lengths and their distributions in the MBL phase. We found that localization lengths extracted from distinct observables do not coincide in general, but all of them remain short throughout the MBL phase. Thus the apparent MBL phase at small system sizes seems stable 
with respect to rare configurations of disorder hosting thermalizing grains; this is consistent with a scenario in which the true MBL critical point occurs at comparable disorder to the numerically observed one. 

The spatial correlation functions and couplings from which we extracted localization lengths share the feature of having log-normal distributions at large separation, with a width that broadens as the separation increases. This feature was noticed in previous work as an approach to a $1/f$ distribution; here, we identify it as a broadening log-normal, a type of behavior that is qualitatively similar to what happens in Anderson localization 
\cite{DMPK1, DMPK2}. The interplay between interaction effects and these broad distributions gives rise to qualitatively modified spectral signatures: both the statistics of local spectral lines and the response to the ``DEER'' spin echo protocol differ qualitatively from naive predictions that ignore the broadening of distributions. We expect similar qualitative modifications for other physical quantities (e.g., post-quench response functions and a.c. conductivity) in which localization lengths appear in the exponent; these consequences will be explored in future work.  

\textit{Note added}.---While this manuscript was being completed, a related work was posted~\cite{paola}, which presented a different algorithm for extracting l-bits and the distribution of localization lengths. 

\section{Acknowledgments}

We thank W. Bialek, T. Can, D. A. Huse, and A. Scardicchio for discussions. D.P. and V.O. also thank B. K. Clark and E. Kapit, respectively, for prior exploratory collaborations that nucleated some of the ideas in this work.
VKV and VO acknowledge support from the NSF DMR Grant No. 1508538 and US-Israel BSF Grant No. 2014265.
SG acknowledges support from NSF Grant No. DMR-1653271.

\section*{Appendix}

\subsection{Further details on Sec. \ref{sec:observables}: 2-spin coupling protocol}
In our work, to expound on Eq. \eqref{eq:H2}, the representation of the L-bit representation is by subsuming multispin interactions into a function that further dresses an effective two-spin model:
\begin{equation}
 \label{eq:lbit2}
 \tilde{H} = E_0 + \sum_i B_i \tau_i^z + \sum_i \sum_{r} J^z_r \tau_i^z \left( \sum_{m=1}^{r-1} \prod_{k=1}^{m}\tau^z_{i+k} \right) \tau^z_{i+r}.
\end{equation}
Here too the interactions are beyond nearest neighbour; with this re-representation, however, we will find a much more systematic variation of the couplings with system size, and thence a better 
definition of localisation regions in space.

In this protocol we are interested in generating the couplings $J^z_r$ as written in Eq. \eqref{eq:lbit2}.
For a given sample and any range $r$, there are $2^{L-2}$ such couplings.

For each state $\psi^{(L-2)}$ in the $2^{L-2}$-dimensional Hilbert space, we take

\begin{equation}
\label{eq:constraint}
\psi_{r_1, r_2} = \psi^{(L-2)} \otimes \{r_1, r_2 \},
\end{equation}

where $r_1, r_2$ denote up or down spins (4 combinations), with the constraint that their site positions $x(r_i)$ on the lattice are given by 

\begin{equation}
 x(r_1) - x(r_2) = r.
\end{equation}

Then for each $\psi^{(L-2)}$ we solve the linear equation 

\begin{equation}
 \begin{bmatrix} +1 & -1 & -1 & +1 \\ +1 & +1 & -1 & -1 \\ +1 & -1 & +1 & -1 \\ +1 & +1 & +1 & +1 \\ \end{bmatrix} \times \left[ \begin{array}{c} E_0 \\ J_1 \\ J_2 \\ \tilde{J}_{12} \end{array} \right] = \left[ \begin{array}{c} E_{\psi_{0,0}} \\ E_{\psi_{0,1}} \\ E_{\psi_{1,0}} \\ E_{\psi_{1,1}} \end{array} \right], 
\end{equation}

where the $E_{\psi_{r1,r2}}$'s are the eigenvalues corresponding to that $l$-bit configuration obtained from Wegner flow. \\
Then $J^z := \tilde{J}_{12}$ gives the renormalized coupling from the four states $\psi_{r_1, r_2}$. 
This is repeated over all $\psi^{(L-2)}$, and $x(r_1), x(r_2)$ (again such that Eq. \eqref{eq:constraint} is satisfied). \\

\begin{figure*}[tthp!]
\centering 
\hspace{-1.65 cm}
\subfloat{{\includegraphics[width=19 cm]{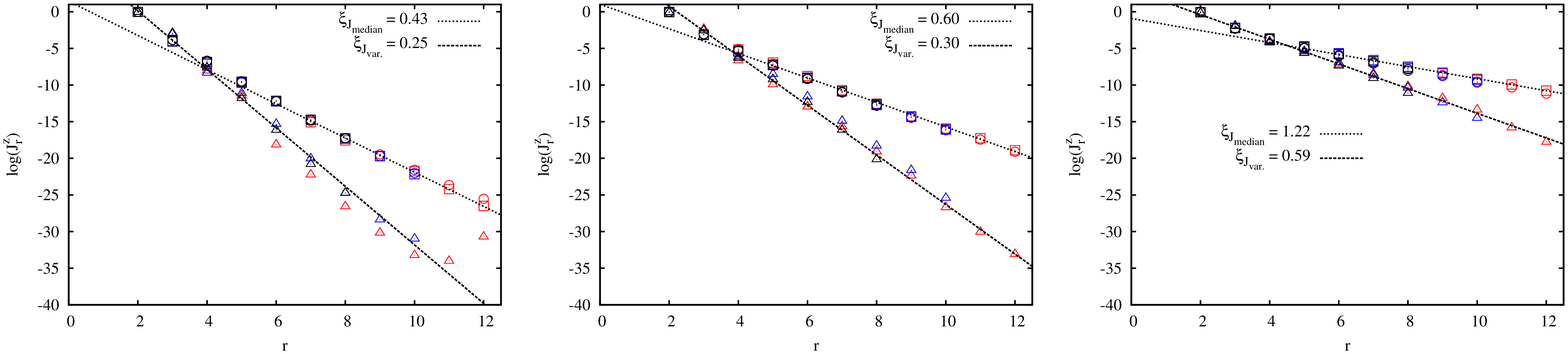}}}%
\\
\hspace{-1.65 cm}
\subfloat{{\includegraphics[width=19 cm]{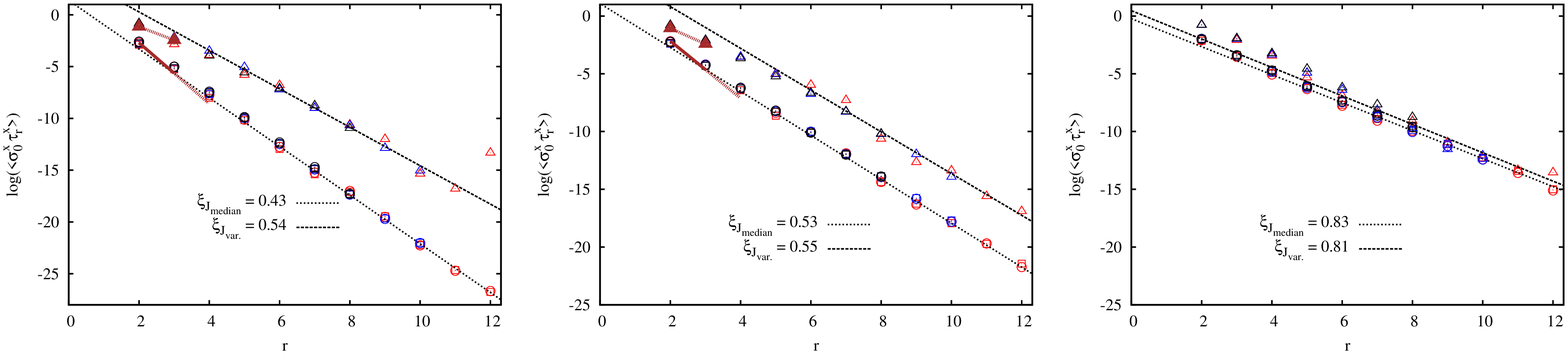}}}%
\\
\hspace{-1.65 cm}
\subfloat{{\includegraphics[width=19 cm]{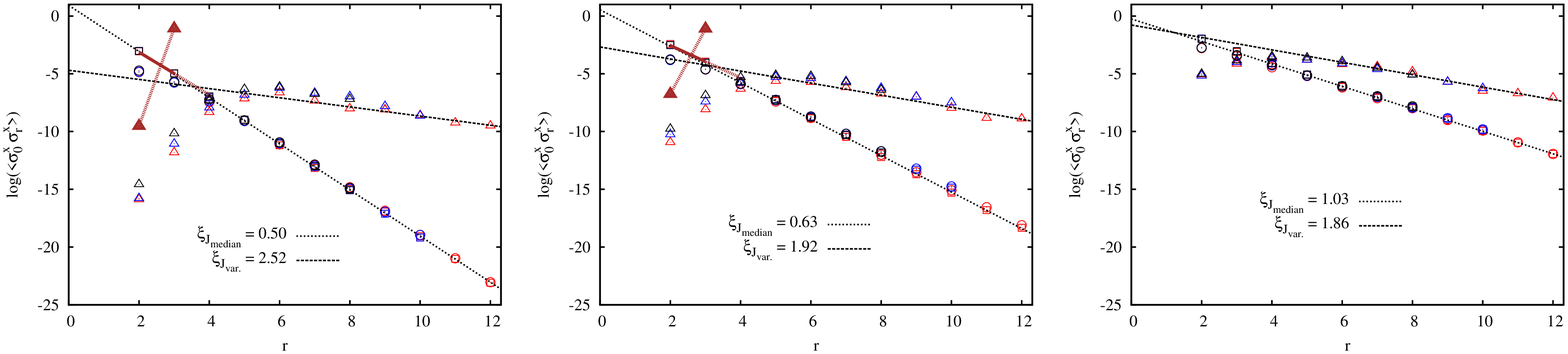}}}%
\caption{DEER decays of couplings for $L=8,10,12$ (black, blue, red) and $2W=25,15,8$ (left to right) using median of $J$ (squares), mean of $\log{J}$ (circles, mostly indistinguishable from squares), 
and variance of $\log{J}$ (triangles). 
A straight line indicates exponential decay, with the two exponential decay lengths from exponential fit of $L=12$ data indicated as above: as disorder increases $\xi$ decreases.
Dotted line are fits to medians, dashed lines are fits to variances, and full brown thick lines are predictions from perturbation theory for the typical values of $\xi$ for $r=2, 3$ 
(the thick brown dot-dashed line shows this perturbation theory prediction $-$ both for log-means and the variances $-$ that is linearly extended to the next site $r=4$ for the median).
\textit{Top row}: decay of $J^z$ couplings in L-bit Hamiltonian. 
\textit{Second row}: decay of $\langle \sigma^x_0 \tau^x_r \rangle$ correlator. Note that the decay lengths are about the same as in top panel for $J^z$ coupling, more so within the localised phase.
The variances are quite well-behaved with about the same decay length scale.
\textit{Third row}: decay of $\langle \sigma^x_0 \sigma^x_r \rangle$ correlator. Note that the decay lengths are about the same as in top panel for $J^z$ coupling, more so away from the localised phase. 
The variances here however have strong finite size effects and are larger than for the median of $J$ and mean of $\log{J}$.
In all rows (where finite-size effects are under control), $\xi$ increases with decreasing $W$ except for fluctuations of mixed operators in middle row i.e. their variances increases with increasing disorder.
}
\label{fig:Jmed2}
\end{figure*}

\begin{figure*}[ttp!]%
    \centering    
\hspace{-1.4cm}    \subfloat[$W=25$]{{\includegraphics[height=3.5cm, width=6.5cm]{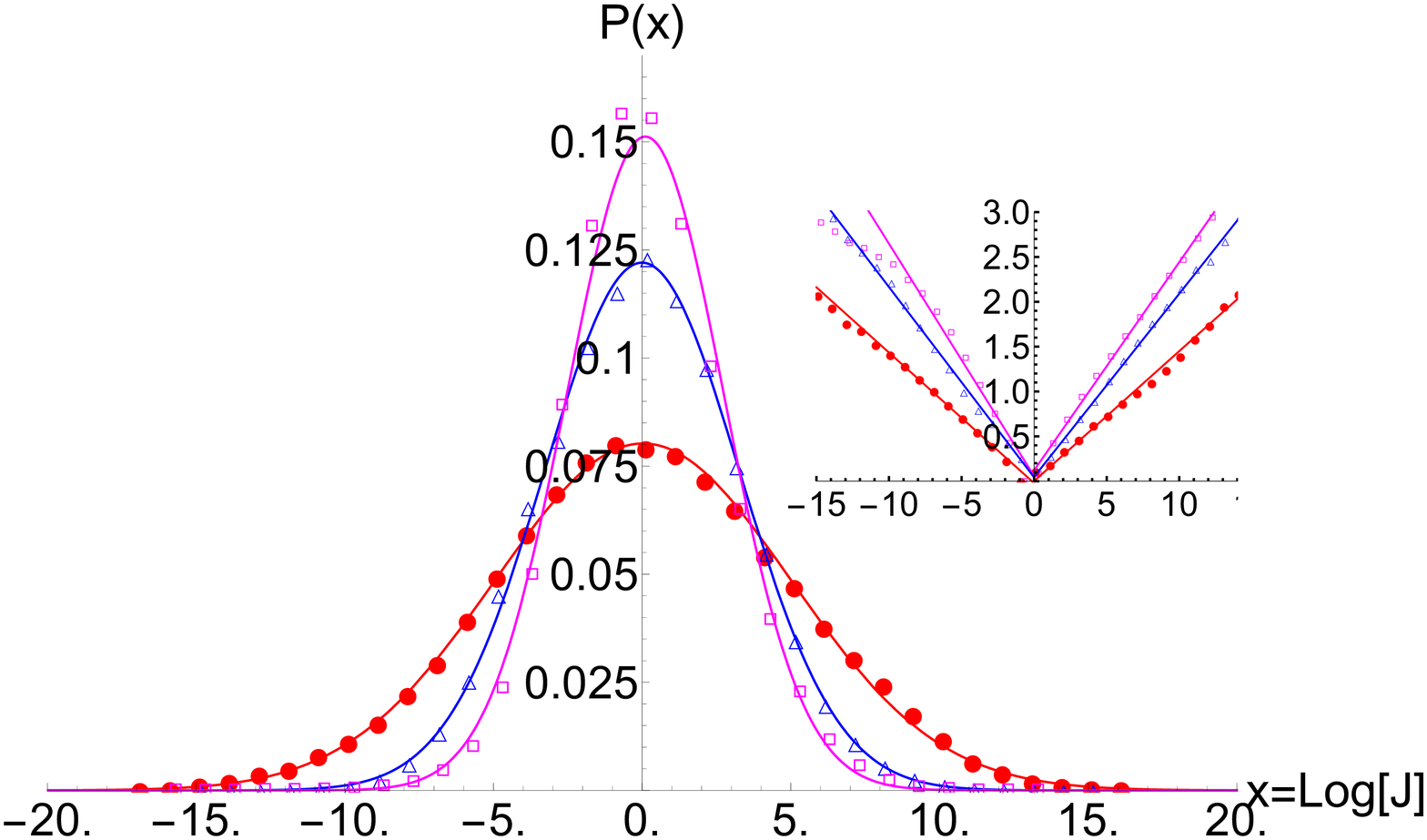} }}%
  \subfloat[$W=15$]{{\includegraphics[height=3.5cm, width=6.5cm]{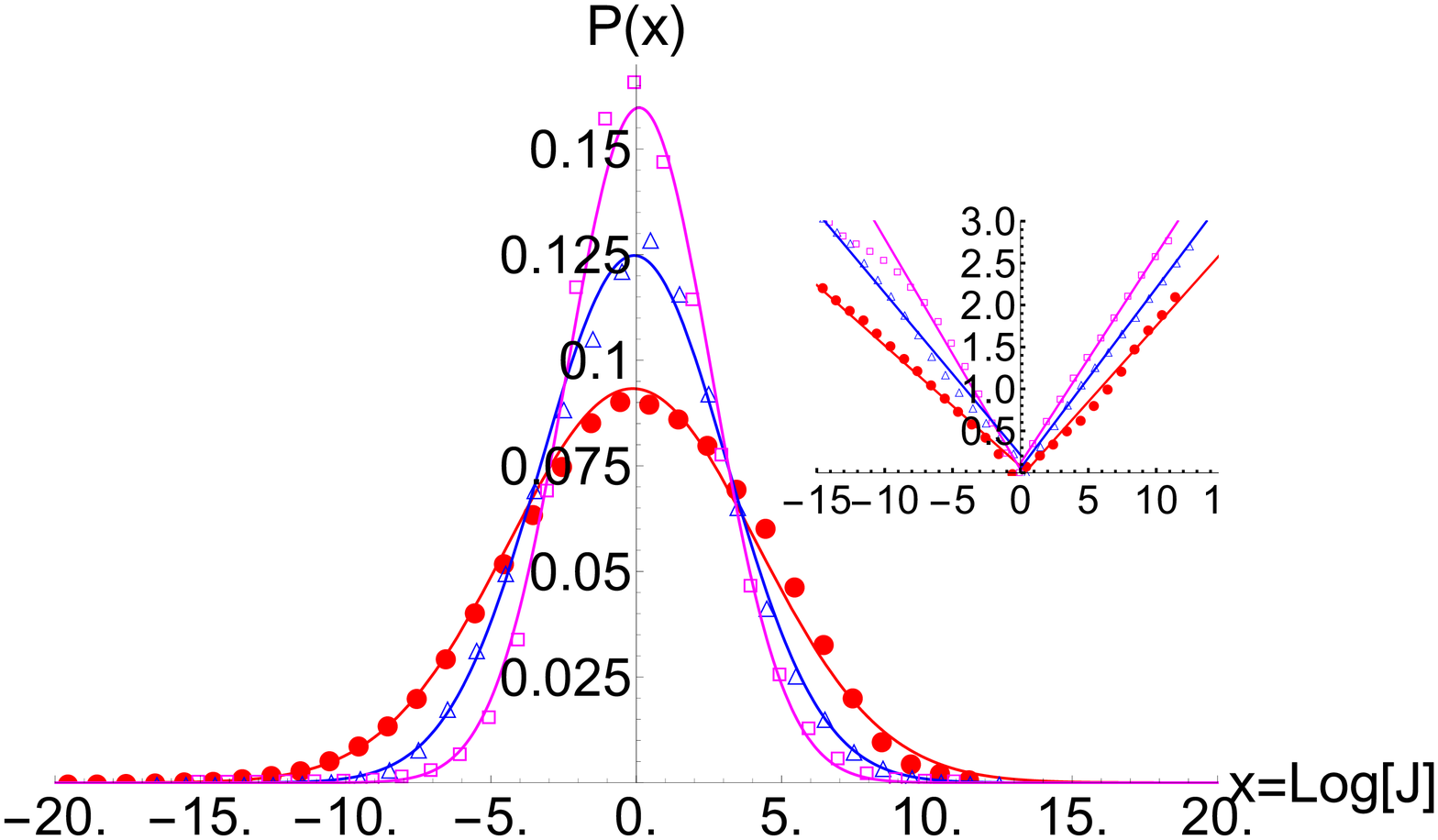} }}%
    \subfloat[$W=8$]{{\includegraphics[height=3.5cm, width=6.cm]{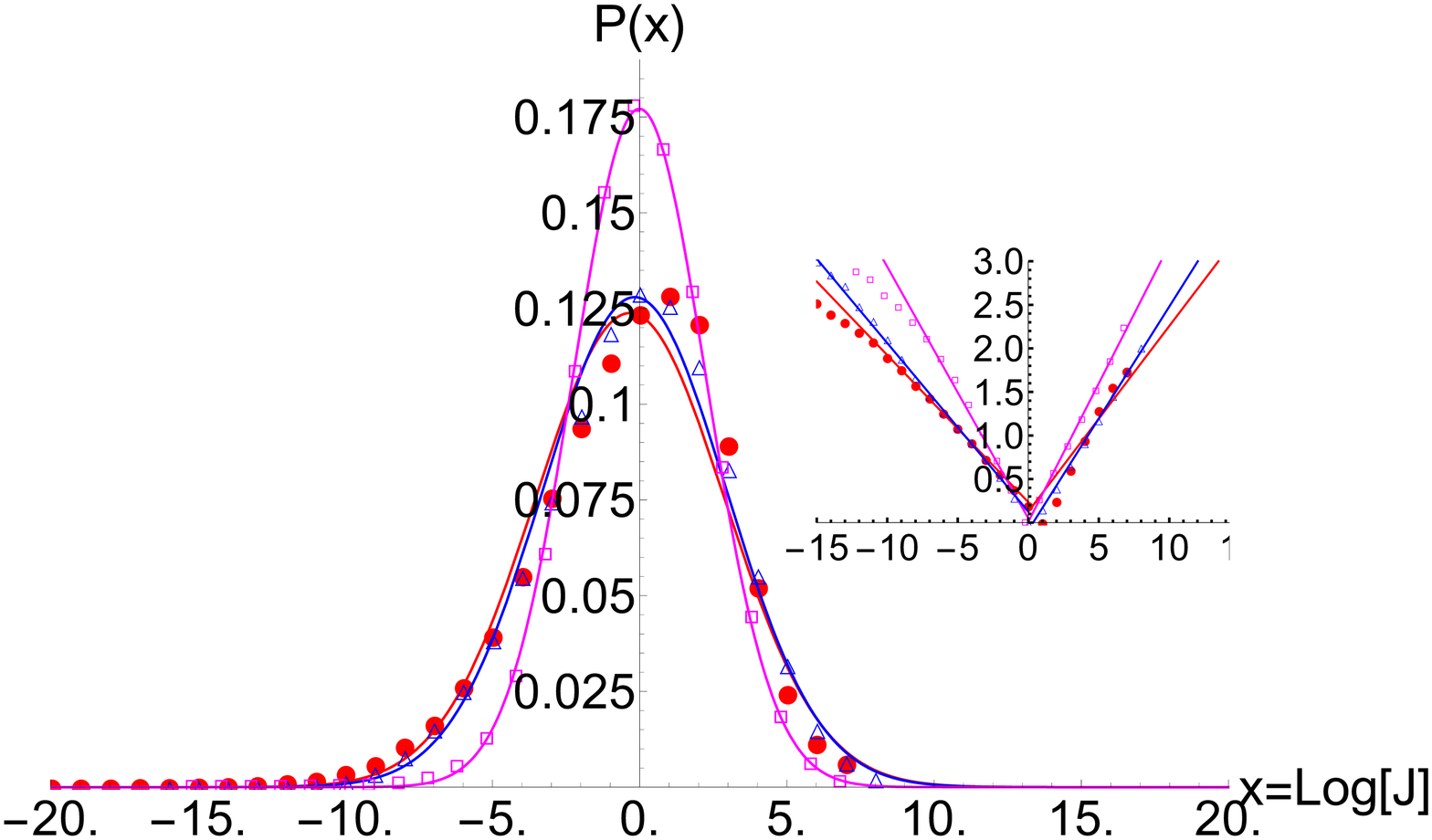} }}%
    \caption{Distributions of logarithms of $J^z, \langle\sigma^x\tau^x\rangle, \langle\sigma^x\sigma^x\rangle$.
   Distribution of logarithm of couplings ($J^z$ (red/circles), $\langle\sigma^x\tau^x\rangle$ (blue/triangles), $\langle\sigma^x\sigma^x\rangle$ (magenta/squares)) for $L=10$ at distance $r=8$, 
and corresponding normal distributions (lines) whose parameters are determined from the data. 
Inset shows square-root of logarithm of y-axis, with lines being fits to straight lines; deviations from straight-lines are indicative of deviations from log-normality of the corresponding coupling 
(strongest for xx coupling).
    }%
    \label{figApp:Jz-vs-xX}%
\end{figure*}
\subsection{Further details on Sec. \ref{sec:observables}: multispin coupling protocol}

In this protocol we are interested in generating the couplings $J^z_r$ as in previous section but with an additional index that denotes the number of coupled $\tau^z$ operators. 
For a given sample and range $r$, there are $2^{r}$ such couplings.

Consider L-bits $0000, 0001, 0010, \ldots$ such that $\tau_i |0\rangle = +1|0\rangle$. Then after diagonalizing:

\begin{equation}
\begin{bmatrix} +1 & +1 & \ldots & +1 & +1 & \ldots \\ -1 & +1 & \ldots & -1 & -1 & \ldots \\ \vdots & \vdots & \vdots\end{bmatrix} \times \left[ \begin{array}{c} E_0 \\ J_1 \\ \vdots \\ J_{12} \\ J_{13} \\ \vdots \end{array} \right] = \left[ \begin{array}{c} E_{0000} \\ E_{0001} \\ \vdots \end{array} \right] 
\end{equation}

The $J_{i}, J_{i,j}, J_{i,j,k} \ldots$ can be obtained by solving above linear equation. \\

The right-hand side are simply eigenvalues from WWF. The left-hand-side matrix is known a priori because of ordering of L-bits in the given sector.

For $r=L$ there are equal number $2^{L-2}$ of $J^z$ and $J^z_{\textrm{ms}}$ couplings; using $L=3, 4$ we can show that these two sets of couplings, written below as a vector, are related by 
\begin{equation}
 \label{eq:DEERvsTEXAS}
 J^z_{\textrm{ms}} = \left(\prod_{i=1}^{L-2} \otimes a \right) J^z,
\end{equation}
where $a = \frac{1}{2}\begin{bmatrix}1 & 1\\ -1 & 1\end{bmatrix}$. So we see that even for the largest range, the two sets of couplings will be different numerically: however their decay tendencies are 
\textit{qualitatively} similar.

\subsection{Spectral trees: mean-field and beyond}

In a general MBL system, the location of the spectral line depends on the global configuration of L-bits, via the interaction terms in Eq.~\eqref{eq:lbit2}. The state of every L-bit $j \neq i$ affects that of L-bit $i$, though the effect falls off exponentially with the separation as $\exp(-|j - i|/\xi)$. The effects of the other spins on the local field at site $i$ can be understood in terms of a spectral ``tree'' with $2^{L - 1}$ nodes at the edge, each corresponding to the effective local field in a many-body eigenstate. 

The structure of this spectral tree was first discussed in Ref. \cite{NGH}, in a treatment that implicitly assumed a unique, sharply defined localization length; we refer to this, below, as a ``mean-field'' treatment.
According to this treatment, when $\xi < 1/(2\ln 2)$, the splittings due to distant lines fall off sufficiently fast that the thermally averaged spectral function does not fill in; instead, it forms a fractal structure with spectral gaps at all frequency scales 
\cite{NGH}. This can be seen as follows: the typical splittings at scale $L$ are $\sim \exp(-L/\xi)$, and those at scale $L + 1$ are $\exp(-(L+1)/\xi)$. When $\exp(-1/\xi) \alt 1/4$, the branches at stage $L+1$ coming from the four different nodes at stage $L$ typically do not cross. The resulting thermally averaged spectral function has gaps at scales $\exp(-n/\xi)$ for all $n$, analogous to a Cantor set. 

This ``mean-field'' treatment can be used to derive the distribution of gaps between adjacent spectral lines in the local, thermally averaged L-bit spectral function. 
A simple version of this argument, valid for $\xi \ll 1$, is as follows. Consider the probability of a given spectral gap $\delta$ exceeding some threshold $\delta_0$, 
i.e., $P(\delta > \delta_0)$; these gaps correspond to splittings at very early stages in the spectral tree, and thus to events at distances $r \alt r_0 \approx \text{const.} - \xi \log \delta_0$. 
The total number of such splittings scales as $P(\delta > \delta_0) \approx 2^{r_0} \sim \delta^{-\xi \log 2}$. Differentiating this to get the probability distribution, we find that 
\begin{equation}
\label{eq:MF-PdB}
P(\delta) \sim \delta^{-1 - \xi \log 2}.
\end{equation}
This analytic prediction agrees with numerical simulations of the mean-field theory (including multiplicative noise of order unity, which is inevitably present because of the randomness of matrix elements). 
While the mean-field theory predicts a continuously varying exponent that is always greater than unity, numerically we find that $P(\delta) \sim 1/\delta$ throughout the MBL phase. 
In the main text, we have ``fixed'' this mean-field picture by including noise in the exponent, so that there is a crossover from Eq.\eqref{eq:MF-PdB} to $1/\delta$ behavior at a given $\delta B$ 
(which depends on the strength of the noise.); see Fig. \ref{fig:SpectralProps}(b).

Beyond mean-field, in our case, the local fields are given by $B_i$ of Eq. \eqref{eq:lbit2}. They are generated by taking the difference in eigenvalues of L-bit states that are flipped at site $i$:
\begin{equation}
 \label{eq:Bdef}
 B_i = E_{\ldots 1_i \ldots} - E_{\ldots 0_i \ldots}; 
\end{equation}
this equation is only true on average, because upon taking the difference on RHS a whole bunch of $J^z$ couplings will also enter into the mix that dress the $B_i$ fields.\\
To be contrete, at first order in perturbation theory of couplings, these are split into $B_i \pm 2\Delta$ for $i$ in the bulk and 
$B_i \pm \Delta$ for $i$ at the boundaries. Already at this level, we see how the level spacing $\delta B$ is proportional to the coupling $J$.
Introducing further spins generates more splittings; this will become clearer if we employ the multispin couplings $J^z_{ijk\ldots}$ for a boundary spin. 
At first order the splitting is $\pm J^z_{12} \approx \Delta$; at next order the upper branch ($B_i + J^z_{12}$) has new splittings of $\pm (J^z_{13} + J^z_{123})$, whereas the lower branch ($B_i - J^z_{12}$) has 
new splittins of $\pm (J^z_{13} - J^z_{123})$, and so on. 
A spectral tree will thus be built up at each site, whose splittings are determined by combinations of $J^z$ couplings.
\subsection{Hamming distance}

The Hamming distance is defined as the number of spin flips required to go from one spin configuration (whether L-bit or p-bit) to another.
For example the states $1011$ and $0111$ have a Hamming distance of two between them. 
Clearly there is some arbitrariness in how we choose from which pair the states to construct the Hamming distance; however following our discussion of local fields and their splittings 
(especially Fig. \ref{fig:SpectralProps}(a),(b)) we see that adjacent pairs of L-bit states (ordered by their local fields) are a good indicator of localisation, i.e. broader the distribution of $P(\delta B)$, 
smaller the typical $\xi$, and stronger the localisation.\\
This immediately implies that adjacent states in a strongly localised system will be connected to each other by fewer number of spin flips than in an ergodic system 
(where the median field splitting is much larger). This expectation is borne out, as displayed in the inset of Fig. \ref{fig:SpectralProps}(d) where we display the mean Hamming distances as a function of 
disorder strength for three system sizes $L=8,10,12$. There appear to be stronger finite size effects as we approach the ergodic phase, which considerably decrease as we wade deeper into the localised phase. \\
Moreover, interestingly, in the main panel of Fig. \ref{fig:SpectralProps}(d) we see that the Hamming distances, sorted over all samples and all states, follow a Poisson distribution as we enter deep into the localised phase.
This means that there is an emergent pair-wise statistical independence of spin-flips among adjacent states as we enter deeper into the localised phase.
The exponential distribution of spin-flips conspicuously breaks down as the disorder is decreased, and the ergodic phase is entered.
The reason for this change in distribution is unclear to us for now but it provids us a clear spectroscopic probe of localisation that is readily amenable to spin-flip experiments. \\
The main upshot of this Appendix is that the Hamming distances clearly demarcate localised vs. ergodic physics, whether through their means or the distributions.\\

\subsection{Perturbation theory}
\begin{figure*}%
    \centering
    \captionsetup{position=top}
    \subfloat[Onsite potentials]{{\includegraphics[width=5.5cm]{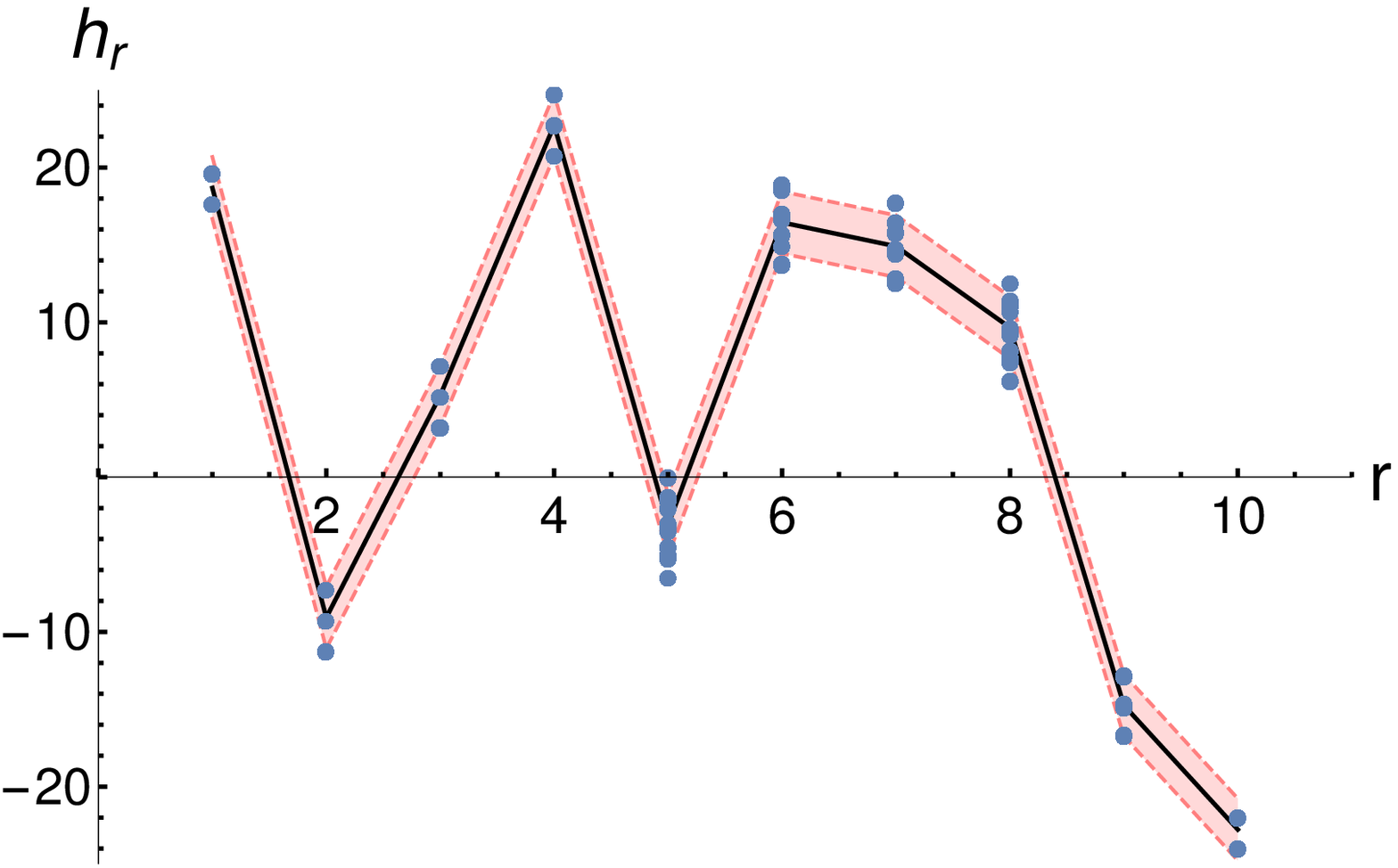} }}%
    \subfloat[Couplings decay]{{\includegraphics[width=5.5cm]{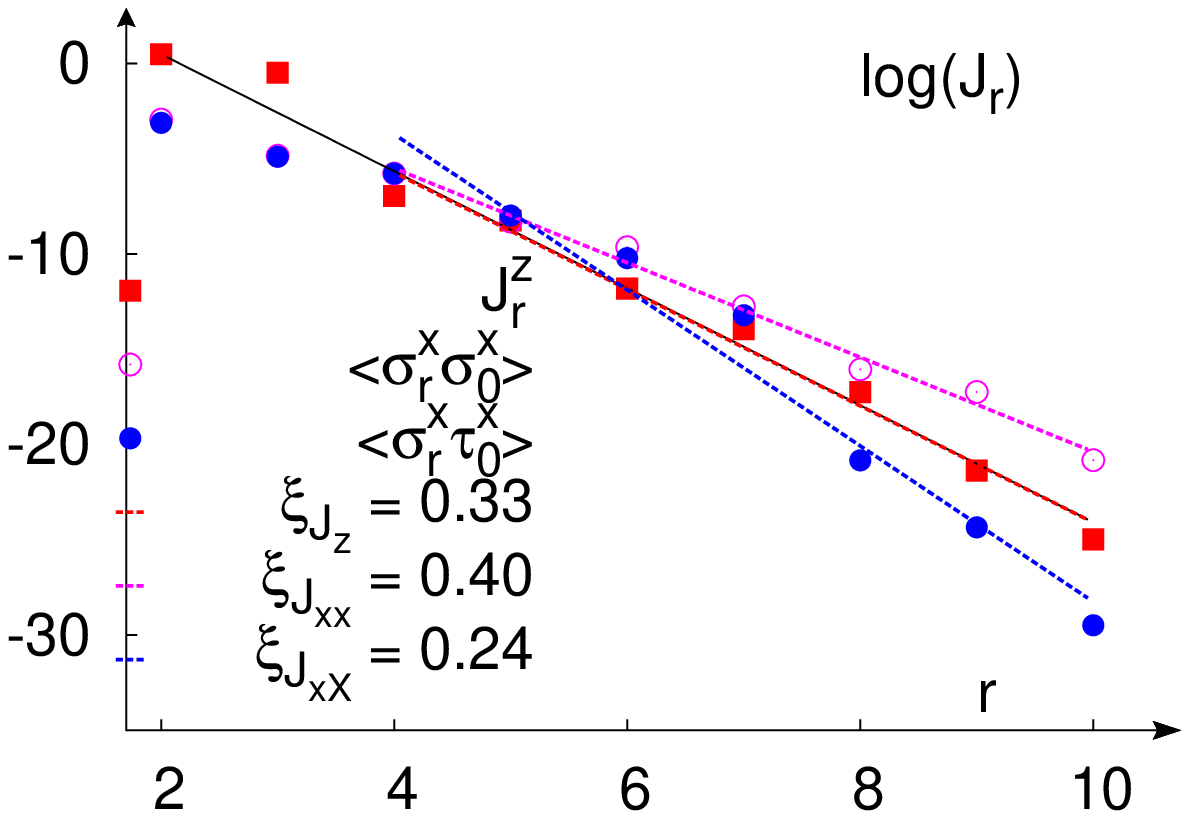} }}%
    \subfloat[Difference of local fields]{{\includegraphics[width=6.0cm]{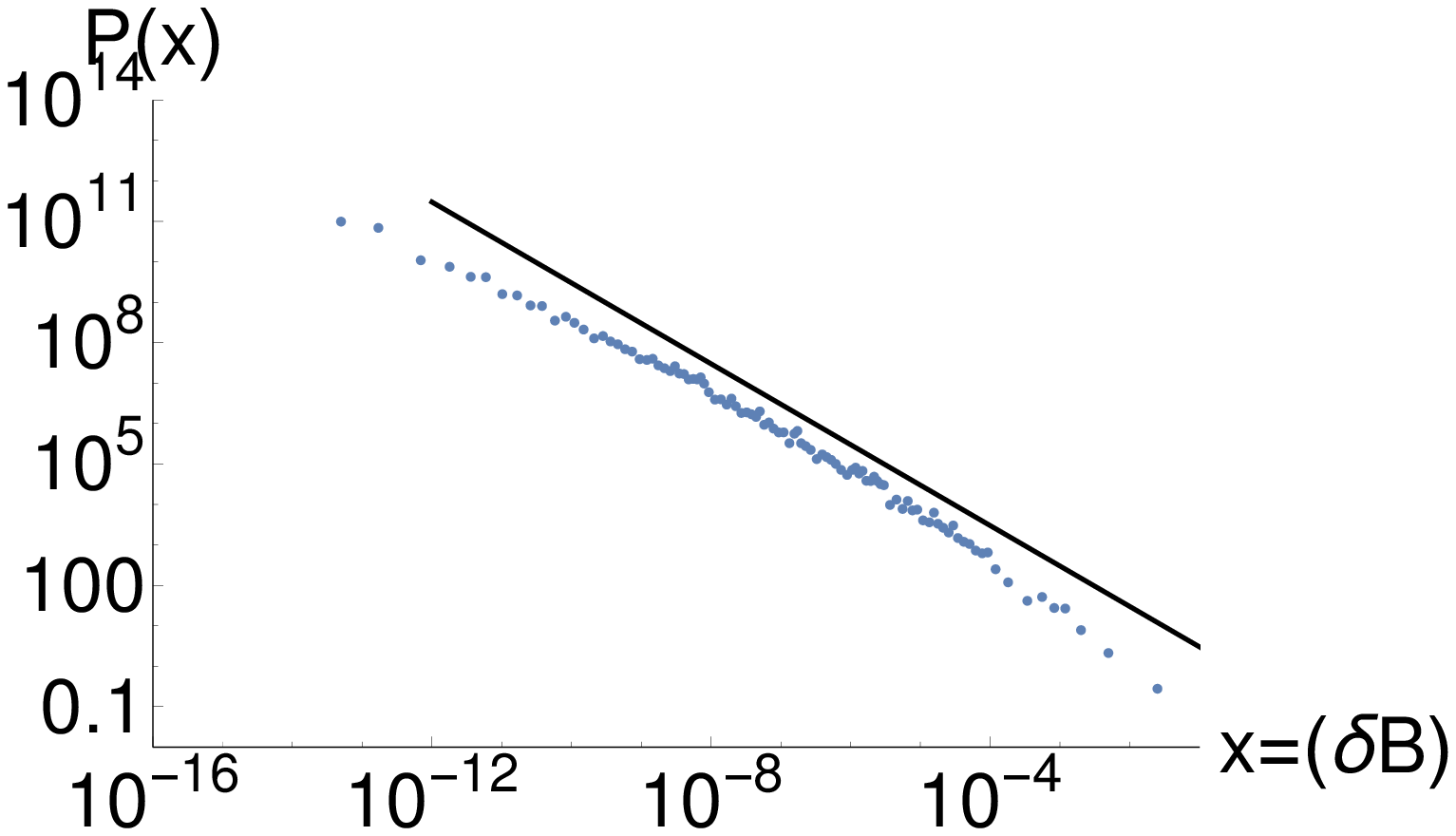} }}%
    \\
    \subfloat{{\includegraphics[width=5.5cm]{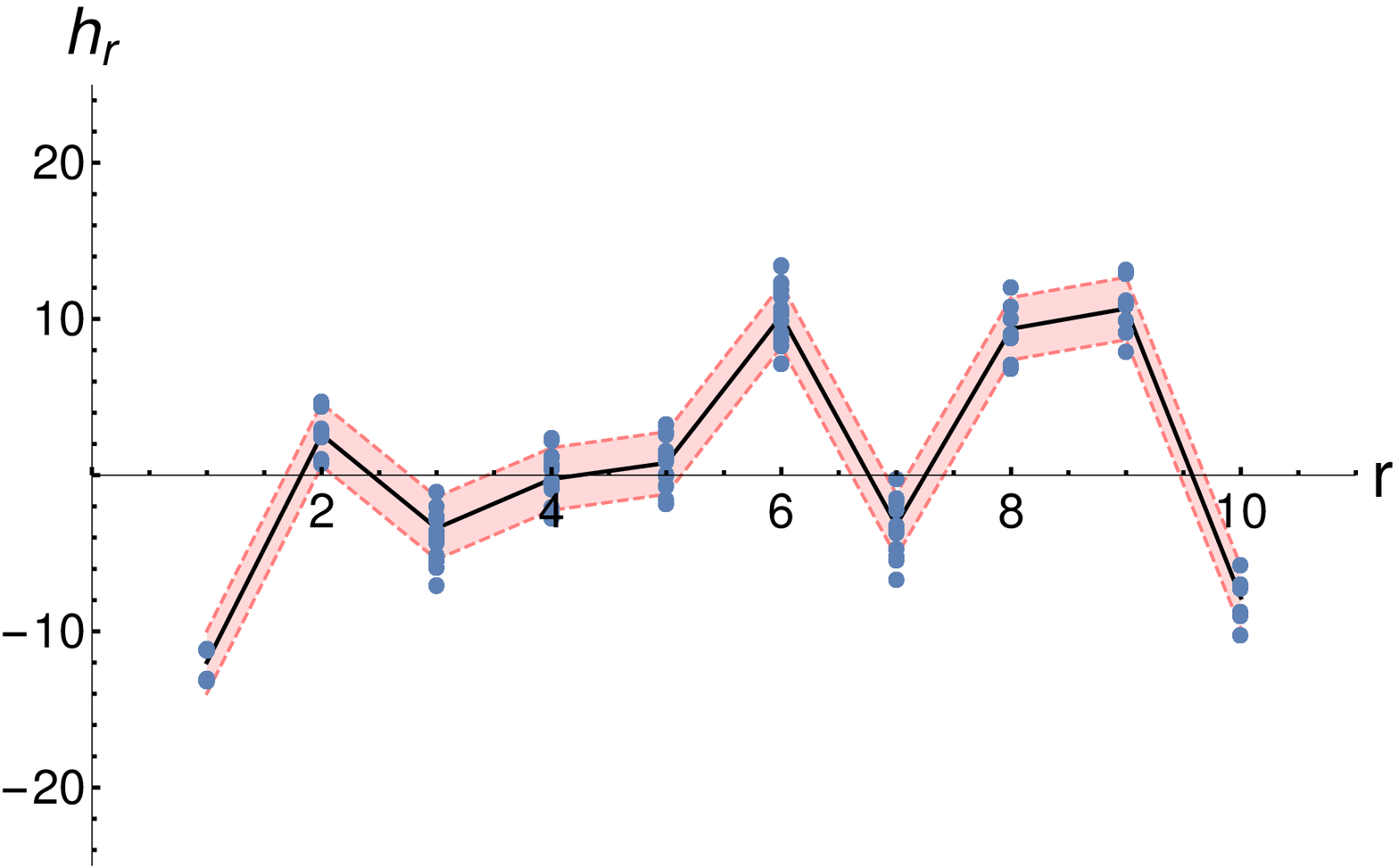} }}%
    \subfloat{{\includegraphics[width=5.5cm]{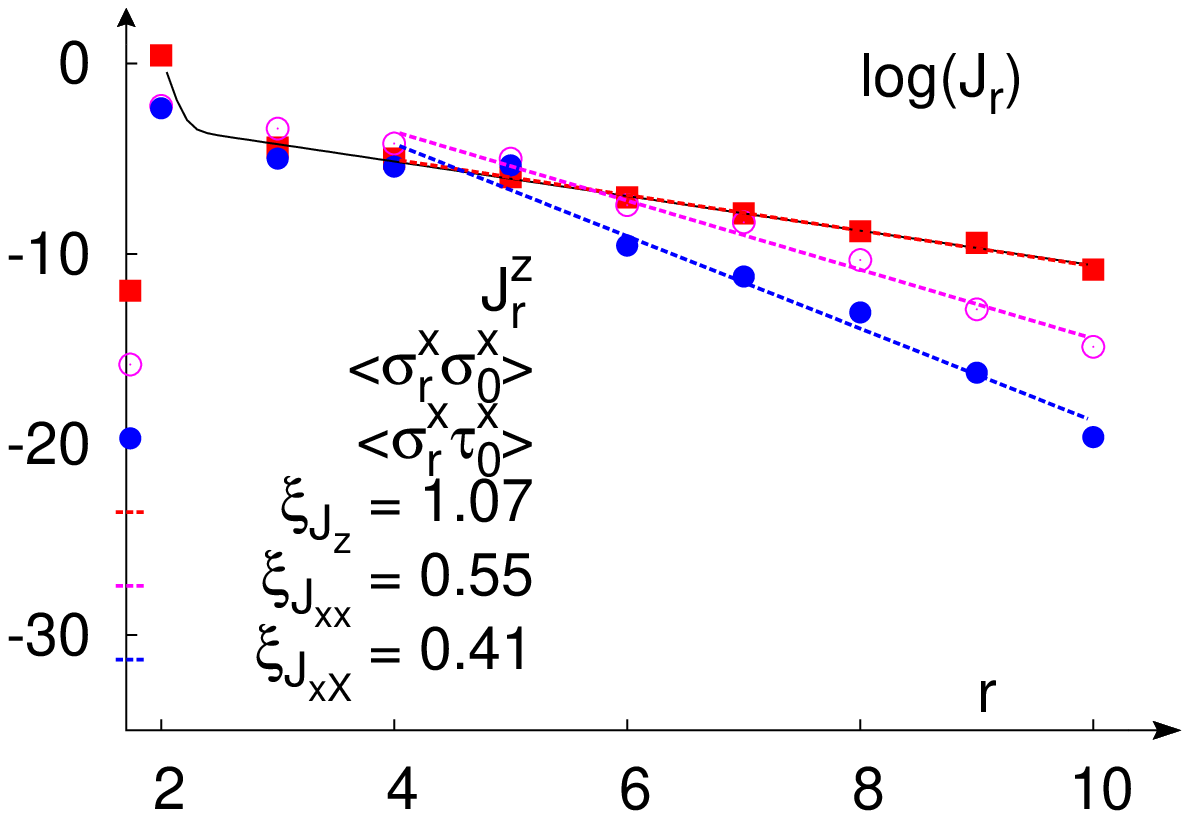} }}%
    \subfloat{{\includegraphics[width=6.0cm]{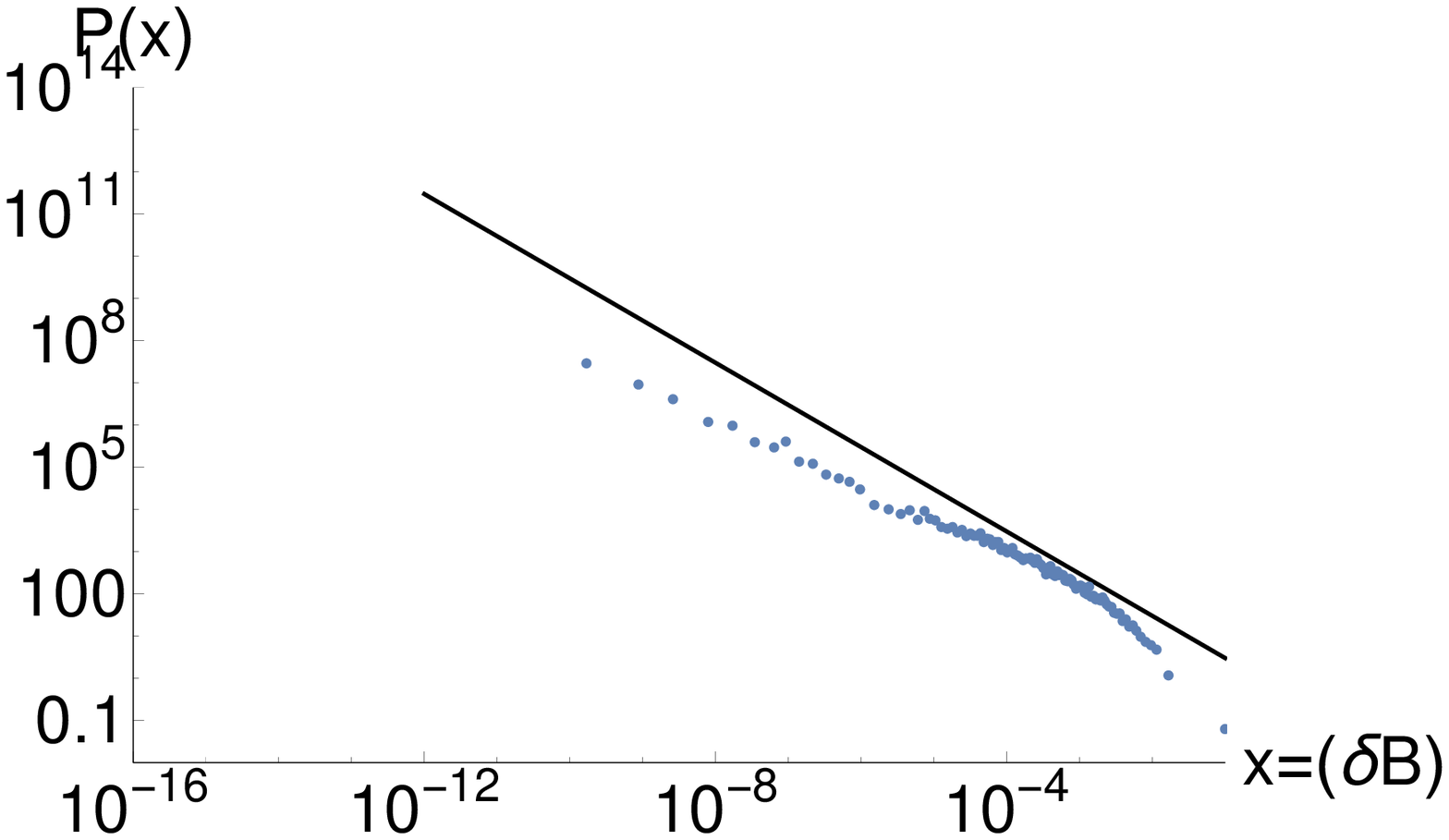} }}%
    \caption{Single sample analysis with small (top row) and large (bottom row) $\xi$ for $W=25$. 
    Column (a) shows the onsitie disorder potentials with the shaded pink area indicating the $\pm 2 \Delta = \pm 2$ area expected for the splitting in local potentials by the interactions $\Delta$, 
    in first order perturbation theory. The blue circles are the local fields obtained from WWF.
    Column (b) shows the decay of the three correlators for the respective sample; the top figure clearly shows a stronger decay than the bottom figure for all three correlators with the single-$\xi$ 
    fits over latter 60\% of data shown as dashed lines of the same colour for each correlator. The thin black full line is a double-$\xi$ fit to $J^z$ couplings over all the data range (see text).
    This stronger decay is reflected in the distribution of local field splittings as shown in column (c), 
    which have a much broader distribution in the strongly localised sample (top) than in the more delocalised sample (bottom). The black line indicates a $1/f$ distribution.
    }%
    \label{fig:SingleSample}%
\end{figure*}
Our starting point is a three site chain with strong onsite potentials, with the Hamiltonian $H_0$ diagonal in these fields and the interaction of strength $\Delta$. 
Using second order perturbation theory in xy terms $H_1 = J_{xy}(\sigma^x_0\sigma^x_1 + \sigma^y_0\sigma^y_1 + \sigma^x_1\sigma^x_2 + \sigma^y_1\sigma^y_2)$ we may perturbatively 
construct the eigenstates up to second order in $J_{xy}$.
Let us denote the difference in local disorder fields between sites to be $\delta_{02}, \delta_{10}, \delta_{12}, \delta_{02}$; with $J_p = 2J_{xy}$, the fields of $\sigma^x \sigma^x$ and $\sigma^x \tau^x$ 
correlators are
\begin{widetext}
 \begin{eqnarray}
  \label{eq: Pertxx}
  \langle \sigma^x_0 \sigma^x_1 \rangle &=& \{ 0, 0, \frac{J_p^3}{4\delta_{02}(\delta_{12} \pm \Delta)^2}, \pm \frac{J_p}{\delta_{10} + \Delta}, \pm \frac{J_p}{\delta_{10} - \Delta}\} \nonumber \\
  \langle \sigma^x_0 \sigma^x_2 \rangle &=& \{ 0, 0, \frac{J_p^2}{2\delta_{02}(\delta_{12} \pm \Delta)}, \frac{J_p^2}{2(\delta_{12}\pm \Delta)(\delta_{10}\pm \Delta)}, -\frac{J_p^2}{2\delta_{02}(\delta_{10} \pm \Delta)} \}; \nonumber \\ 
 \langle \sigma^x_0 \tau^x_1 \rangle &=& \{ \frac{J_p}{2(\delta_{10} \pm \Delta)}, -\frac{J_p}{2(\delta_{10} \pm \Delta)}, \frac{J_p}{2(\delta_{10} \pm \Delta)}, -\frac{J_p}{2(\delta_{10} \pm \Delta)} \} \nonumber \\
 \langle \sigma^x_0 \tau^x_2 \rangle &=& \{ 0, 0, \frac{J_p^2}{4\delta_{02}(\delta_{12} \pm \Delta)}, 2 \times \frac{J_p^2}{4(\delta_{10}\pm \Delta)}\left[\frac{1}{\delta_{12} \mp \Delta} - \frac{1}{\delta_{02}} \right] \}. 
 \end{eqnarray}
 \end{widetext}
 Note that it is vital to include the $J_p^3$ contribution in the first line lest the zeros dominate whilst taking the typical values of the above fields: this is because the denominators of these $J_p^3$ terms can skew 
 these otherwise parametrically smaller fields to be comparable to the other terms proportional to $J_p$. (This may be checked for a given small $L$ system comparing numerics and the above expressions.)\\
Taking $J_{xy} = J^z = 1/W$, with $W \gg 1$, we see that the typical value of $-(\log{\frac{\langle \sigma^x_0 \sigma^x_2 \rangle}{\langle \sigma^x_0 \sigma^x_1 \rangle}})^{-1} \approx 0.54, 0.73$ and 
 $-(\log{\frac{\langle \sigma^x_0 \tau^x_2 \rangle}{\langle \sigma^x_0 \tau^x_1 \rangle}})^{-1} \approx 0.33, 0.41$ from these perturbative treatments for $W=25, 15$ respectively; 
 these perturbation theory predictions are plotted in Fig. \ref{fig:Jz-vs-xX}(a) and shown as dot-dashed thick brown lines, both for the means and variances.
while the analytical predictions for typical $\sigma^x \sigma^x$ correlators are in reasonable agreement with the numerical values for $W \gg 1$, the agreement for the $\sigma^x \tau^x$ correlators is much less impressive.
This is presumably due to nonperturbative effects in the latter.
Moreover we see that the initial upturn (strong-finite size effect) in the variance for $\sigma^x \sigma^x$ correlators at small $r$ is nicely captured by the perturbation theory, as is the absence of this upturn in the $\sigma^x \tau^x$ 
correlators; these are indicated by the thick brown triangles joined by the thick dot-dashed lines.

Therefore after averaging this inverse localisation lengths we find generically that 
\begin{equation}
\label{eq:xxVsxX}
 \xi^{-1}_{\textrm{xx}} < \xi^{-1}_{\textrm{xX}}, 
\end{equation}
thereby making the $\langle \sigma^x_0 \tau^x_2 \rangle$ correlator decay faster than 
$\langle \sigma^x_0 \sigma^x_2 \rangle$.
We have seen this to be true whether averaged all samples (Fig. \ref{fig:pxi}) or in single samples (Fig. \ref{fig:SingleSample}).\\
Moreover we see from perturbation theory that as $\Delta \rightarrow 0$, (i) the two localisation lengths for $\sigma^x \sigma^x$ and $\sigma^x \tau^x$ will be different;
(ii) $\xi \sim \Delta^{\kappa}$, with $\kappa \approx 0.25 < 1$. The linear dependence on $\Delta$ (i.e. $\kappa = 1$) is satisfied only when $\Delta/W \approx 0.1$.

\subsection{Variance measures}
Unlike the medians or means, the variances over states and samples do not commute i.e. it depends whether we perform the variance measurements over all states and samples, 
or first mean over states and then variance over samples (intersample variance), or first variance over states and then mean over samples (intrasample variance).
Therefore in principle from the variances alone there will be three length scales $\xi_{\textrm{var.}}, \xi_{\textrm{var. mean}}, \xi_{\textrm{mean var.}}$, 
which correspond to the above three ways respectively.
In the main text we presented only the first of the three. However it is clear that the full variance over samples and states is simply the sum of the other two. \\
Moreover we see that it is not always the case that intersample fluctuations dominate; this seems only true for $\langle \sigma^x_0  \sigma^x_r \rangle$ and $\langle \sigma^x_0  \tau^x_r  \rangle$ correlators 
but not for $J^z$ couplings. While the latter two do not have strong transients at small $r$, the p-bit correlator does. 
This is explained by the perturbative structure of fields, already seen in Eq. \eqref{eq: Pertxx} where there is a larger proportion of exact zero values in the correlators at a given order.
Indeed we find that the results from perturbation theory at small distances captures the numerical results for variances too, 
qualitatively and quantitatively, discriminating between intersample and intrasample fluctuations.\\
Finally in all three measures of variances, we find that a linearly growing variance leads to a broad distribution of these couplings as well, whether in the intrasample or intersample measure 
i.e. the presence of broad distributions is a persistent effect.

\bibliographystyle{unsrt}
\bibliography{Ref}

\end{document}